\documentclass[11pt]{article}
\usepackage[usenames,dvipsnames,svgnames,table]{xcolor}
\usepackage{xparse}
\usepackage{tikz}
\usetikzlibrary{positioning,patterns,matrix,arrows,calc}
\usetikzlibrary{decorations.markings}
\usetikzlibrary{patterns}
\usetikzlibrary{snakes}
\usetikzlibrary{shapes}
\usetikzlibrary{trees}
\usetikzlibrary{decorations.pathmorphing}
\usepackage{amsfonts}
\usepackage{amssymb}
\usepackage{amscd}
\usepackage{amsmath}
\usepackage{epsfig}
\usepackage{graphicx, subfigure}
\usepackage{latexsym}
\usepackage[font=small,labelfont=bf]{caption} 
\usepackage{verbatim}
\usepackage{tipa}
\usepackage{amsthm,amssymb}

\newlength{\dinwidth}
\newlength{\dinmargin}
\newcommand{\be}{\begin{equation}}
\newcommand{\ee}{\end{equation}}
\newcommand{\ba}{\begin{eqnarray}}
\newcommand{\ea}{\end{eqnarray}}

\newtheorem{definition}{Definition}
\newtheorem{theorem}{Theorem}

\newtheorem{remark}{Remark}
\newtheorem{lemma}{Lemma}
\newtheorem{example}{Example}

\def \m{{\rho}}

\begin{document}

\title{Feynman diagrams, ribbon graphs, and topological recursion of Eynard-Orantin}
\author{K. Gopala Krishna$^1$, Patrick Labelle$^2$, and Vasilisa Shramchenko$^3$}

\date{}

\maketitle

\footnotetext[1]{Department of mathematics and statistics, Concordia University, E-mail: {\tt gopala.krishna@concordia.ca}}

\footnotetext[2]{Champlain Regional College, Lennoxville campus, Sherbrooke, Quebec, Canada. E-mail: {\tt plabelle@crc-lennox.qc.ca}}

\footnotetext[3]{Department of mathematics, University of
Sherbrooke, 2500, boul. de l'Universit\'e,  J1K 2R1 Sherbrooke, Quebec, Canada. E-mail: {\tt  Vasilisa.Shramchenko@Usherbrooke.ca}}

 \begin{abstract}
 We consider two seemingly unrelated problems, the calculation of the WKB expansion of the harmonic oscillator wave functions  and the counting the number of  Feynman diagrams in QED or in many-body physics and  show that their solutions are both encoded in a single enumerative problem,  the calculation of the  number of certain types of  ribbon graphs. In turn, the numbers of such ribbon graphs as a function of the number of their  vertices and edges can be determined recursively through the application of the topological recursion of Eynard-Orantin  to the algebraic curve encoded in the Schr\"odinger equation of the harmonic oscillator. We show how the numbers of these ribbon graphs can be written down in closed form for any given number of vertices and edges. We use these numbers to obtain a formula for the number of $N$-rooted ribbon graphs with $e$ edges, which is the same as the number of Feynman diagrams for $2N$-point function with $e+1-N$ loops.

%We show that the WKB expansion of the wavefunctions of the harmonic oscillator encode the solution to an enumerative problem: the counting of QED-type  Feynman diagrams with or without Furry's theorem imposed.  Each term of the WKB expansion is associated to diagrams with a certain genus  and number of external lines. This gives  a topological interpretation to the WKB expansion, whose terms are  and is reminiscent 
 \end{abstract}

\tableofcontents

\section{Introduction}

\noindent In this paper we consider two seemingly  completely unrelated physical problems : calculating the WKB approximation of the  harmonic oscillator wave functions and counting the number of Feynman diagrams  in many-body physics, which correspond to the diagrams one would obtain from QED if Furry's theorem were not applicable.   As we will show, the answers to both problems can actually be obtained from a single formalism,  the topological recursion of Eynard and Orantin,  applied to an algebraic curve of genus zero encoded in a natural way in the Schr\"odinger equation for the harmonic oscillator.

\noindent This leads to the somewhat surprising result that  calculating the  WKB expansion of the  harmonic oscillator wave functions  is related  to solving the  enumerative problem of counting the number of Feynman diagrams in  QED or in many-body physics.   This also gives a link 
between a topological problem and  the WKB expansion of the harmonic oscillator.

%Beginning with the seminal work of t'Hooft \cite{thooft} on the use of matrix integrals to study the large N limit of Yang-Mills, much work has been devoted to applications of matrix models to various topics in quantum physics, for example in string theory, in 2D quantum gravity or in topological field theory (see for example \cite{Witten}, \cite{dif}, \cite{Kont} . In parallel, matrix models have been applied to enumeration problems of various types of combinatorial objects \cite{Itzykson}.  

\noindent Inspired by the loop equations of matrix models,  the topological recursion  of Eynard and Orantin - that we will now simply refer to as topological recursion -  offers an algorithm to generate an infinite hierarchy of multi-differentials  starting from  a spectral curve and, at least in the case of genus zero curves, no other input. The nomenclature of spectral curves originates from the first applications where the curves were associated to the spectra of matrix models but the formalism has by now been applied to much more  general situations, where curves may have no connection to any matrix model.  What has generated much interest in this formalism is that, with judicious choices of spectral curves, these multi-differentials solve a large number of enumerative problems, in the sense that the coefficients of well-chosen  series expansions of these multi-differentials count various quantities of topological or combinatorial interest. For example, topological recursion has been used to reproduce the Witten-Kontsevich intersection numbers of moduli spaces of curves, the Weil-Petersson volumes of moduli spaces of hyperbolic surfaces, the stationary  Gromov-Witten invariants of $\mathbb{P}_1$, simple Hurwitz numbers and their generalizations, Gromov-Witten invariants of toric Calabi-Yau threefolds and much more (see for example \cite{Eynard_ICM} for an overview of some applications).

\noindent In addition to the construction of generating functions for enumerative problems, the multi-differentials may be used to construct a so-called wave function defined as the exponential of a sum of integrals of the multi-differentials. One then defines {\it quantum curves} as differential operators annihilating this wave function. For a large class of genus zero curves, the quantum curves such defined are obtained from the corresponding spectral curves by making the simple substitutions $x \rightarrow x, y \rightarrow \hbar d/dx$  (note that $y$ does not correspond to the canonical momentum of quantum mechanics) and by  making a specific choice of ordering  when ordering  ambiguities are present, see \cite{BouchardEynard}.   In those instances, the construction of the wave function using the differentials of topological recursion can be shown to coincide with the usual WKB expansion. 
 When the dependence on $y$ is at most quadratic,  the application of the  quantum curve  on the wave function to give  zero takes the form of the Schr\"odinger equation. 
  The situation is more delicate for higher genera, see for example \cite{bouchardstudents} and references therein.

\noindent In the present paper, we focus on one of the simplest  genus zero spectral  curves $y^2 = x^2 - c^2$, for which the quantum curve  is simply $\hbar^2 d^2/dx^2- x^2 + c^2$ and topological recursion reproduces the WKB expansion of the  quantum  harmonic oscillator \cite{BouchardEynard,HO}.

\noindent However, we notice, following \cite{DumitrescuMulaseSafnukSorkin} and \cite{MotohicoCatalan},  that in addition to giving the WKB expansion of the wave function,   the same differentials can be used to compute certain numbers 
$C_{g,n} (\mu_1, \dots, \mu_n)$ which are  objects of study in combinatorics, see \cite{Tutte1,Tutte2,WalshLehman1,WalshLehman2,WalshLehman3}. They
represent the numbers of maps (graphs drawn on a compact orientable surface in a way that each face is a topological disc) on a genus $g$ surface having $n$ ordered vertices of respective degrees $\mu_i$  and such that at each vertex one of the half-edges incident to this vertex is marked.  
 It has been known for some  time that using these numbers,  one can count the number of rooted maps (that is maps with a distinguished half-edge) of genus $g$ with $e$ edges \cite{WalshLehman1}. It has been recently realized, see \cite{Feynman} and \cite{3authors}, that the rooted maps are in one-to-one correspondence with  the Feynman diagrams of  the two-point function of  a charged scalar field interacting with a neutral scalar field through a cubic term $\phi^\dagger A \phi $. These diagrams also correspond to the electron propagator in QED if Furry's theorem is not applied or in many-body physics,  including tadpoles. In the rest of the paper we will follow the example of \cite{cvitanovic} and refer to our diagrams as QED diagrams and call the two types of particles   electrons and photons, but one must keep in mind that our diagrams will include electron loops connected to arbitrary numbers of photons. We will also give formulas for the number of QED diagrams with Furry's theorem enforced  in Section 7. 
 
\noindent The connection between ribbon graphs and Feynman diagrams has been studied and exploited since the seminal work of 't Hooft \cite {TH} allowing to see ribbon graphs as Feynman diagrams through the matrix model approach. However, the correspondence between Feynman diagrams and ribbon graphs thus obtained is very different from ours. In \cite{A} and \cite{B} an approach similar to ours was used for the vacuum diagrams of the QED theory we consider here.

\noindent  In \cite{Feynman} it is shown that in order to extend this correspondence  between the  QED Feynman diagrams of the electron propagator  and rooted maps to   Feynman diagrams containing an arbitrary number of external electrons, one is naturally led to introducing the notion of $N$-rooted maps, that is maps with $N$ distinguished half edges. 
 In this paper, we show that it is also possible to use the  $C_{g,n} (\mu_1, \dots, \mu_n)$ to compute  the number of $N$-rooted maps with a given number of edges.
 % and that these maps are in bijection with the Feynman diagrams of a quantum field theory of a charged scalar field interacting with a neutral scalar field through a simple cubic interaction.
 
\noindent The application of topological recursion to the harmonic oscillator spectral curve therefore provides a curious bridge between three apparently unrelated problems: one mathematical, the enumeration of a certain type of maps,  and the other two physical,  the computation of the WKB expansion of the wave function and of the number of Feynman diagrams in  QED.

\noindent The paper is organized as follows.  In Section 2 we review the topological recursion of Eynard and Orantin as applied to the harmonic oscillator curve and we show how  to compute  differentials   $W_{g,n} (\mu_1, \dots, \mu_n)$
used to obtain the coefficients $C_{g,n} (\mu_1, \dots, \mu_n)$.
%can be extracted from the differentials generated by topological recursion. 
In Section 3, we show  how the differentials obtained in the previous section are used to obtain the WKB approximation of the harmonic oscillator wave functions.
In Section 4, we  introduce ribbon graphs and maps and describe  how their enumeration is encoded in a set of coefficients $C_{g,n}  (\mu_1, \dots, \mu_n)$.  In Section 5, we show how closed form formulas for  the coefficients  $C_{g,n}  (\mu_1, \dots, \mu_n)$ can be obtained from the differentials $W_{g,n} (\mu_1, \dots, \mu_n)$ generated by topological recursion. In Section 6 we present the formula giving the number of $N$-rooted maps in terms of the $C_{g,n} (\mu_1, \dots, \mu_n)$ and  explain how they can be used to  count QED diagrams. Section 7 shows the expression for the first non trivial WKB correction to the harmonic oscillator in terms of certain coefficients $C_{g,n}  (\mu_1, \dots, \mu_n)$.   In the   appendix we provide  additional  explicit expressions for some $C_{g,n}$.

\bigskip

\bigskip

\noindent {\bf Acknowledgements.} 
K.G. wishes to thank Dmitry Korotkin, Marco Bertola as well as the staff and members at Concordia University for the support extended during his stay.  P.L. gratefully acknowledges the support from  the Fonds  de recherche du Qu\'ebec - Nature et technologies (FRQNT)  via a grant from the Programme de recherches pour les chercheurs de coll\`ege and to the STAR research cluster of  Bishop's University.  V.S. is grateful for the support from the Natural Sciences and Engineering Research Council of Canada through a Discovery grant as well as from the University of Sherbrooke. P.L. and V.S.  thank the  Max Planck Institute for Mathematics in Bonn, where this work was initiated, for hospitality and a perfect working environment.

   \tikzset{
    photonloop/.style={decorate, decoration={snake,amplitude=1.5pt,segment length=4 pt},draw=black},
        photon/.style={decorate, decoration={snake},draw=black},
    electron/.style={draw=black, postaction={decorate},
        decoration={markings,mark=at position .75 with {\arrow[draw=black,scale=2]{>}}}},
         electronf/.style={draw=black, postaction={decorate},
        decoration={markings,mark=at position .15 with {\arrow[draw=black,scale=2]{>}}}},
              electronff/.style={draw=black, postaction={decorate},
        decoration={markings,mark=at position .25 with {\arrow[draw=black,scale=2]{>}}}},
          electronfff/.style={draw=black, postaction={decorate},
        decoration={markings,mark=at position .85 with {\arrow[draw=black,scale=2]{>}}}},
          electronmiddle/.style={draw=black, postaction={decorate},
        decoration={markings,mark=at position .55 with {\arrow[draw=black,scale=2]{>}}}},
    positron/.style={draw=black, postaction={decorate},
        decoration={markings,mark=at position .55 with {\arrow[draw=black,scale=2]{<}}}},
    gluon/.style={decorate, draw=magenta,
        decoration={coil,amplitude=4pt, segment length=5pt}},
        fermion/.style={draw=black, postaction={decorate},decoration={markings,mark=at position .55 with {\arrow[draw=black,scale=2]{>}}}},
  vertex/.style={draw,shape=circle,fill=black,minimum size=3pt,inner sep=0pt}
}

\NewDocumentCommand\semiloop{O{black}mmmO{}O{above}}
{
\draw[#1] let \p1 = ($(#3)-(#2)$) in (#3) arc (#4:({#4+180}):({0.5*veclen(\x1,\y1)})node[midway, #6] {#5};)
}

\NewDocumentCommand\semiloopt{O{black}mmmO{}O{above}}
{
\draw[#1] let \p1 = ($(#3)-(#2)$) in (#3) arc (#4:({#4+245}):({0.5*veclen(\x1,\y1)})node[midway, #6] {#5};)
}

\NewDocumentCommand\semilooptt{O{black}mmmO{}O{above}}
{
\draw[#1] let \p1 = ($(#3)-(#2)$) in (#3) arc (#4:({#4+285}):({0.3*veclen(\x1,\y1)})node[midway, #6] {#5};)
}

\NewDocumentCommand\semiloopq{O{black}mmmO{}O{above}}
{
\draw[#1] let \p1 = ($(#3)-(#2)$) in (#3) arc (#4:({#4+90}):({0.3*veclen(\x1,\y1)})node[midway, #6] {#5};)
}
\NewDocumentCommand\semiloopf{O{black}mmmO{}O{above}}
{
\draw[#1] let \p1 = ($(#3)-(#2)$) in (#3) arc (#4:({#4+360}):({0.3*veclen(\x1,\y1)})node[midway, #6] {#5};)
}

\NewDocumentCommand\semiloopqq{O{black}mmmO{}O{above}}
{
\draw[#1] let \p1 = ($(#3)-(#2)$) in (#3) arc (#4:({#4+85}):({0.3*veclen(\x1,\y1)})node[midway, #6] {#5};)
}

\section{Topological recursion for the harmonic oscillator curve}

\noindent The Eynard-Orantin topological recursion is a procedure which constructs an infinite hierarchy of multi-differentials defined on a given algebraic curve.   Let us assume that the curve $\mathcal C$ defined by a polynomial $P(x,y)=0$ is of genus zero and the ramified covering $x:\mathcal C\to \mathbb CP^1$ has only simple ramification points. The original definition for any curve with simple ramification is given in \cite{EO} and \cite{BouchardEynard} gives the recursion formula adapted to an arbitrary ramification. 

\noindent Under our assumptions, the Eynard-Orantin topological recursion takes the following form : 
\begin{multline}
\label{recursion}
W_{g,n}(p_1,\dots, p_n) =
\\
= \sum_{i=1}^m \underset{q=A_i}{\rm res} K(q,p_1)\left(  W_{g-1,n+1}(q,q^*, p_2, \dots, p_n) 
+ \sum_{\substack{g_1+g_2=g\\ I\sqcup J =\{2,\dots, n\}}}^{{\rm no }\;(0,1)} W_{g_1,|I|+1}(q,p_I)W_{g_2,|J|+1}(q^*,p_J)
\right)\,,
\end{multline}
where

\begin{itemize}
\item $q, p_i\in\mathcal C$ are points on the genus zero curve $\mathcal C$;
\item $A_i$ are simple ramification points of the covering $x:\mathcal C\to \mathbb CP^1$;
\item $q^*$ denotes the image of the local Galois involution of a point $q\in\mathcal C$ lying in a neighbourhood of a simple ramification point $A_i$ (that is the points $q$ and $q^*$ belong to the two sheets meeting at $A_i$ and project to the same value of $x$ on $\mathbb CP^1$, the base of the covering); 
\item the second sum  excludes the value $(0,1)$  for $(g_1,k)$ and $(g_2,k)$; 
\item the initial ``unstable'' differentials $W_{0,1}$ and $W_{0,2}$ are given by (being points on a genus zero curve $p, p_1,p_2\in\mathcal C$  can be thought of as complex numbers):
\begin{equation}
\label{w02}
W_{0,1}(p)=y(p)dx(p)\,,\qquad W_{0,2}(p_1,p_2) = \frac{dp_1 dp_2}{(p_1-p_2)^2}\,\,;
\end{equation}
\item the recursion kernel $K$ is defined by 
\begin{equation}
\label{K}
K(q,p_1)=\frac{1}{2}\frac{\int_q^{q^*}W_{0,2}(\xi,p_1)}{W_{0,1}(q)-W_{0,1}(q^*)}\,\,.
\end{equation}
\end{itemize}

\noindent This is a recursion with respect to the number $2g-2+n>0.$
The obtained multi-differentials $W_{g,n}$ are invariant under arbitrary permutation of their arguments $p_1,\dots, p_n$, have poles at the ramification points $A_i$ with respect to each of the arguments and no other singularities, see \cite{EO}. \\

\vskip 0.8cm
\noindent Let us consider a family of {\it harmonic oscillator curves} parameterized by $c\in\mathbb C$, that is a family of algebraic curves given by the equation
\begin{equation}
\label{HOfamily}
y^2=x^2-c^2\;.
\end{equation}
Away from the point $(x,y)=(\infty, \infty)$ on the curve \eqref{HOfamily}, one can choose the following local parameter: 
\begin{equation*}
z=\sqrt{\frac{x-c}{x+c}}\,\,. 
\end{equation*}
Then we have 
\begin{equation}
\label{localparameter}
 x=-c\frac{z^2+1}{z^2-1}\,, \qquad y = \epsilon\frac{2cz}{z^2-1}\,,\qquad dx=\frac{4czdz}{(z^2-1)^2}\,,
\end{equation}
where $\epsilon=\pm1$ is an arbitrary choice of sign reflecting the arbitrariness in the choice of the sign of the square root defining the parameter $z$. The simple ramification points of the curve are at $z=0, \infty$ and the Galois involution $*$ acts by $z^*=-z$ which corresponds to $(x,y)^*=(x,-y)$. 

\noindent With this data definitions \eqref{recursion}-\eqref{K} give (the superscript $H$ reflects the fact that the quantities are calculated using the harmonic oscillator family of curves \eqref{HOfamily}):
\begin{equation}
\label{HOinitial}
W^{H}_{0,1}(z) = ydx = \epsilon\frac{8c^2z^2\,dz}{(z^2-1)^3}, \qquad\qquad
W^{H}_{0,2}(z_1,z_2) = \frac{dz_1dz_2}{(z_1-z_2)^2}
\end{equation}
and
\begin{equation}
\label{HOkernel}
K^{H}(z,z_1)=\frac{\epsilon}{16 c^2}\frac{(z^2-1)^3 \,dz_1}{(z^2-z_1^2)z \,dz}\;.
\end{equation}
%

%\begin{example} 
\noindent Now, computing differentials $W_{g,n}$ for the harmonic oscillator curve reduces to plugging these data into recursion formula \eqref{recursion} and computing residues of some rational expressions. Here are a few examples taken from \cite{HO}. 

\noindent The first generation, with $2g-2+n=1$:
\begin{eqnarray*}
&&W_{0,3}^H(z_1,z_2,z_3)=\epsilon \frac{dz_1dz_2dz_3}{2^3c^2}\left(1-\frac{1}{z_1^2z_2^2z_3^2} \right); 
\nonumber
\\
&& W_{1,1}^H(z_1)= \epsilon\frac{(z_1^2-1)^3\,dz_1}{2^6\,c^2\, z_1^4}\,.
\nonumber
\end{eqnarray*}
\noindent The second generation, with $2g-2+n=2$:
\begin{equation*}
 W_{0,4}^H (z_1,z_2,z_3,z_4)= \frac{dz_1dz_2dz_3dz_4}{2^6c^4}
\left(
\frac{3}{ (z_1z_2z_3z_4)^2} \sum_{i=1}^4\frac{1}{z_i^2}  - \frac{9}{(z_1z_2z_3z_4)^2}  -\sum_{i<j}\frac{1}{(z_i z_j)^2} - 9 + 3\sum_{i=1}^4 z_i^2
\right);
\end{equation*}
\begin{multline}
W_{1,2}^H(z_1,z_2) = \frac{dz_1dz_2}{2^9 c^4} \left( \frac{5}{(z_1 z_2)^2} \sum_{i=1}^2 \frac{1}{z_i^4}  +\frac{3}{ (z_1 z_2)^4} 
-\frac{18}{(z_1 z_2)^2}  \sum_{i=1}^2 \frac{1}{z_i^2} +\frac{27}{(z_1 z_2)^2}-4  \sum_{i=1}^2 \frac{1}{z_i^2 } \right.
\\
\left. + 27
-18  \sum_{i=1}^2 z_i^2  + 5 \sum_{i=1}^2 z_i^4
+3 (z_1 z_2)^2
 \right). \label{w12}
\end{multline}

\section{WKB wave function}
\noindent We will briefly review how the WKB expansion for the harmonic oscillator can be obtained from the $W^H_{g,n}$, see \cite{DumitrescuMulaseSafnukSorkin}, \cite{HO} for more details. 

\noindent Let us write the one-dimensional time independent Schr\"odinger equation  in the form
\begin{equation*}
- \hbar^2 \frac{d^2 }{dx^2} \Psi(x,\hbar) + 2 M V(x) \, \Psi(x,\hbar) =  2M E \,  \Psi(x, \hbar )  \, .\end{equation*}
Let us define 
\begin{equation*}  2 M  V(x) = f(x) 
\end{equation*}
and
\begin{equation*}
 c^2 = 2  M  E  ,
\end{equation*}
so that Schr\"odinger's equation becomes
\begin{equation}
\biggl( \hbar^2  \,  \frac{d^2 }{dx^2} - f(x) + c^2 \biggr)  \,  \Psi(x,\hbar) =  0. \label{schroe}
\end{equation}
\noindent This differential operator is called the quantum curve and corresponds to the algebraic (spectral) curve of the equation $y^2 -f(x) + c^2=0\,. $ For a large class of curves, see \cite{BouchardEynard},  the WKB wave functions can be calculated from the $W_{g,n}$ of the corresponding spectral curve. Here we outline this calculation for the example of the harmonic oscillator curves \eqref{HOfamily}, that is specializing to the case  $V(x) =  M \omega^2 x^2/2$ and setting $M \omega =1$. 
We first introduce the functions $F_{g,n}^H$ with $2g-2+n>0$ defined by integrating $W_{g,n}^H$ as follows
\begin{equation}
\label{FgnHO}
F^H_{g,n}(z_1,\dots, z_n) =\frac{1}{2^n} \int^{z_1}_{-z_1} \dots \int^{z_n}_{-z_n} W^H_{g,n}(z_1' , \dots,  z_n'). 
\end{equation}
\noindent For example, one finds
\begin{eqnarray*}
&&F^H_{0,3}(z_1,z_2,z_3)= \frac{\epsilon}{2^3c^2}\left(z_1z_2z_3+\frac{1}{z_1z_2z_3} \right);
\nonumber
\\
&&F^H_{1,1}(z_1)=\frac{\epsilon}{2^6\,c^2} \left( \frac{z_1^3}{3} -3z_1 -\frac{3}{z_1} + \frac{1}{3z_1^3}\right).
\nonumber
\\
&&F^H_{0,4}(z_1,z_2,z_3,z_4)=
 \frac{1}{2^6 \, c^4}
\left(
\frac{1}{ z_1 z_2 z_3 z_4 } \sum_{i=1}^4\frac{1}{z_i^2}  - \frac{9}{z_1 z_2 z_3 z_4}  -\sum_{\substack{i<j, k<l\\ i \neq k \neq l\neq j}}
\frac{z_k z_l}{z_i z_j} - 9 \,  z_1 z_2 z_3 z_4  + z_1 z_2 z_3 z_4 \sum_{i=1}^4 z_i^2
\right); \nonumber 
\\
&&F^H_{1,2}(z_1,z_2)= 
 \frac{1}{2^9 \, c^4}
 \left( \frac{1}{z_1 z_2} \sum_{i=1}^2 \frac{1}{z_i^4} 
  +\frac{1}{3 z_1^3 z_2^3} 
-\frac{6}{z_1 z_2}  \sum_{i=1}^2 \frac{1}{z_i^2} 
+\frac{27}{z_1 z_2} + 4 \frac{z_1}{z_2} + 4 \frac{z_2}{z_1}
\right. \nonumber 
\\ && \quad \quad \quad \quad \quad \quad  \quad \quad \quad \quad \quad \quad \quad 
\left. + 27 z_1 z_2
-6 z_1 z_2  \sum_{i=1}^2 z_i^2  + z_1 z_2  \sum_{i=1}^2 z_i^4
+\frac{1}{3} z_1^3 z_2^3
 \right).
\end{eqnarray*}

\noindent  Then the WKB expansion of the wave function satisfying Eq.(\ref{schroe}) is given by 
\begin{equation}
\Psi(x, \hbar) = {\rm exp} \sum_{m=0}^\infty \hbar^{m-1}S_m(x) \label{wkbb}
\end{equation}
with
\begin{eqnarray}
&&S_0(x) = -\frac{1}{2}\int^{(x,y)}_{(x,-y)} y\,dx\;,
\nonumber
\\
&&S_1(x) = -\frac{1}{2}{\rm log}\,y \,,
\nonumber
\\
&&S_m(x) = \sum_{2g+n-2=m-1} \frac{1}{n!} F_{g,n}^{H}(z,z, \ldots z)  \,, \qquad m\geq 2, \label{keysn}
\end{eqnarray}
where in the last equation $z= \sqrt{(x-c)/(x+c)}$.
 Thus we obtain in the region $x>c \, $,
\begin{eqnarray*} 
S_0(x) &=& -  \frac{\epsilon \,  c^2}{2} \ln \bigl(  \sqrt{x^2-c^2}  + x \bigr)+\frac{\epsilon \,  x}{2} \sqrt{x^2-c^2}  ,
\\ S_1(x) &=&   - \frac{1}{2}  \ln \bigl(\sqrt{x^2-c^2} \bigr) \;,
\\ S_2(x) &=&  \frac{1}{3!} F_{0,3}^H(z,z,z) + F_{1,1}^H(z) 
\\
&=& \frac{\epsilon}{24c^2}  \frac{6c^2x-x^3}{(x^2-c^2)^{3/2}}\;,
\\ S_3(x) &=&    \frac{1}{4!} F_{0,4}^H(z,z,z,z) + \frac{1}{2} F_{1,2}^H(z,z)  \\
&=& \frac{3c^6+9 c^4 x^2 - 3 c^2 x^4 + x^6 }{32 \, c^4(x^2-c^2)^3}. 
\end{eqnarray*}
\noindent Using these expressions in Eq.(\ref{wkbb}) reproduces the WKB expansion of the harmonic oscillator wave functions if we set $\epsilon=-1$.

\section{Ribbon graphs and maps}
\label{sect_graphs}

Initiated by the work of Tutte \cite{Tutte1, Tutte2}, enumeration of maps, that is graphs embedded in a compact orientable surface in a way that each face is a topological disc, received much attention in the last century, see \cite{WalshLehman1, WalshLehman2, WalshLehman3, JacksonVisentin, Bender, Bender2, ArquesBeraud}. More recently this topic was extended to counting maps on non-orientable surfaces, see \cite{ArquesGiorgetti, Chapuy, Chapuy2} and references therein. 

\noindent Central to this paper is a notion of a connected ribbon graph, which is equivalent to that of a map. Intuitively, a ribbon graph is a graph whose vertices are replaced by small discs and edges are replaced by ribbons. This can be done in a canonical way once the cyclic order of  edges is fixed at every vertex. 

\noindent In \cite{WalshLehman1}, a closed form expression was given for the number of ribbon graphs with $e$ edges  and with one half edge marked that can be drawn on a genus $g$ surface (as will be defined more precisely below). In \cite{Feynman} this result was rederived using quantum field theory  and  generalized to an arbitrary number of marked half edges. 
%The relevance of these results to the present work lies in the fact %that these marked ribbon graphs are in bijection with the %Feynman diagrams of many-body physics. 
Before explaining this connection in more details, let us define more precisely the quantities we will be using.

\begin{definition}
A \emph{ribbon graph}, or simply \emph{graph}, is a data $\Gamma = (H, \alpha,\sigma)$ consisting of  a set of half-edges $H = \{1,\dots, 2e\}$ with a positive integer $e$ and two permutations $\alpha, \sigma \in S_{2e}$ on the set of half-edges such that
	\begin{itemize}
	\item $\alpha$ is a fixed point free involution,
%	\item $\sigma$ 
	\item the subgroup of $S_{2e}$ generated by $\alpha$ and $\sigma$ acts transitively on $H$. 
	\end{itemize}
\end{definition}

\noindent The involution $\alpha$  is a set of transpositions each of which pairs two half-edges that form an edge. Cycles of the permutation $\sigma$ correspond to vertices of the ribbon graph $\Gamma$; each cycle gives the ordering of half-edges at the corresponding vertex. Cycles of the permutation $\sigma^{-1}\circ\alpha$ correspond to faces of $\Gamma$. The condition of transitivity of the group $\langle \sigma, \alpha \rangle$ on the set of half-edges ensures the connectedness of the graph $\Gamma$. 

\begin{example}
For the graph in Figure \ref{Fig_torus}, we have $H=\{1,2,3,4\}\,,$  $\alpha = (12)(34)\,,$ $\sigma = (1324)\,,$ and  $\sigma^{-1}\alpha = (1324)$, so there is one face.

\begin{figure}[htb]
\centering
\includegraphics[width=5cm]{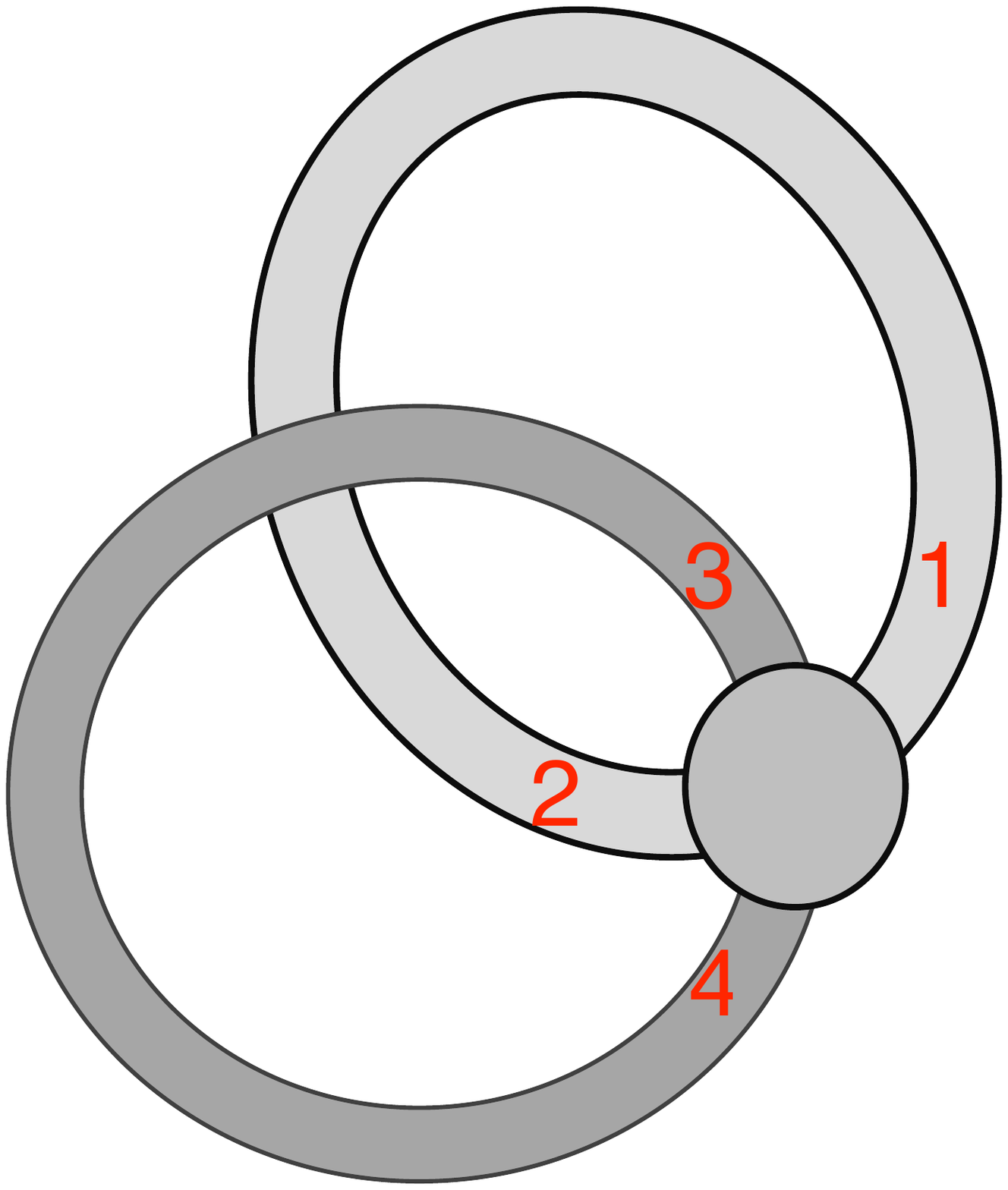}
\caption{A genus one ribbon graph.}
\label{Fig_torus}
\end{figure}

\end{example}

\noindent A ribbon graph defines a connected compact orientable surface. This surface is reconstructed by gluing discs to the faces of the ribbon graph. Let us denote by $n$ the number of vertices of a ribbon graph $\Gamma = (H, \alpha,\sigma)$, that is the number of cycles in the permutation $\sigma$. Denote by $f$ the number of faces, that is the number of cycles in the permutation $\sigma^{-1}\circ\alpha$. Then the genus of the surface corresponding to $\Gamma$ is defined through the Euler characteristic of the graph: 
\begin{equation}
n-e+f=2-2g\,. \label{euler}
\end{equation}
\noindent The genus of the surface is called the genus of the ribbon graph.

\begin{definition}
\noindent An isomorphism between two ribbon graphs $\Gamma = (H, \alpha,\sigma)$ and $\Gamma' = (H, \alpha',\sigma')$ is a permutation $\psi\in S_{2e}$, that is $\psi:H\to H$, such that $\alpha'\circ \psi = \psi\circ \alpha$ and $\sigma'\circ \psi = \psi\circ \sigma$.
\end{definition}

 \noindent That is two graphs are isomorphic if the data of one of them can be obtained from another by relabeling $\psi:H\to H$ of the half-edges. Two isomorphic ribbon graphs are identified.  

\begin{definition}
\noindent For a given graph $\Gamma= (H, \alpha,\sigma)$, an automorphism is a permutation $\psi:H\to H$, such that $\alpha\circ \psi = \psi\circ \alpha$ and $\sigma\circ \psi = \psi\circ \sigma$. The group of automorphisms of $\Gamma$ is denoted by ${\rm Aut}(\Gamma)$. 
\end{definition}

\noindent In other words, an automorphism of a ribbon graph is a relabeling of its half-edges that does not change the permutations $\alpha$ and $\sigma$. 

\begin{example} 
The graph in Figure \ref{Fig_torus} has the following automorphism $\psi=(1423)$, meaning that in the new labeling the half-edge $1$ is labeled by $4$, the half-edge $4$ is labeled by $2$, the half-edge $2$ is labeled by $3$.  The group of automorphisms of this graph is $\mathbb Z_4\,.$
\end{example} 

\begin{definition}
\noindent A \emph{marked graph} is a ribbon graph  whose vertices are labeled with consecutive integers starting at $1$. We call an \emph{unmarked graph} a ribbon graph whose vertices are not labeled. 
\end{definition}

\begin{definition}
A \emph{marked isomorphism} between two marked ribbon graphs is an isomorphism of the corresponding unmarked graphs which acts trivially on the set of vertices, that is on the set of cycles of $\sigma$.
\end{definition}

\noindent  Let us denote by ${\Gamma_v}$ a marked graph whose underlying unmarked graph is $\Gamma$. For a given marked graph ${\Gamma}_v$, the group of automorphisms of $\Gamma$ which fix every vertex is the group of {\it marked automorphisms} of ${\Gamma_v}$ denoted by ${\rm Aut}_v({\Gamma}_v)$. 

\begin{example}
For the ribbon graph $\Gamma$ given by $H=\{1,2,3,4\}\,,\,\alpha=(12)(34)\,,$ and $\sigma=(13)(24)$, we have ${\rm Aut}_v({\Gamma}_v) = \mathbb Z_2\,.$ That is there are two marked automorphisms - the identity and the exchanging of the two edges: $\psi = (13)(24)\,.$ However, if we do not impose that the automorphism fixes the vertices, there is an additional automorphism exchanging the two vertices: $\psi=(12)(34)$ and we have ${\rm Aut}(\Gamma) = \mathbb Z_2\times \mathbb Z_2\,.$
\end{example}

\begin{lemma}
\label{lemma_vertical}
Let $\Gamma_v$ be a marked graph with more than two vertices. Then $|{\rm Aut}_v(\Gamma_v)|=1.$
\end{lemma}

\noindent Given that the group ${\rm Aut}_v({\Gamma}_v)$ does not depend on a particular marking of the underlying graph $\Gamma$, we also write ${\rm Aut}_v({\Gamma})$ for the automorphisms of $\Gamma$ that fix the set of vertices pointwise.

\subsection{The numbers $C_{g,n}$}
\label{sect_Cgn}

\noindent Let us denote by $G_{g,n}(\mu_1, \dots, \mu_n)$ the set of distinct marked graphs of genus $g$ with $n$ vertices for which the vertex labeled $i$ has $\mu_i$ half-edges incident to it. We say that the vertex has {\it degree} $\mu_i$. Since we only consider connected graphs, it is assumed that the degrees $\mu_i$ are positive integers for $n>1$. If $n=1$ the set $G_{0,1}(0)$ consists of one graph which is a single point. 

\begin{figure}[htb]
\centering
\includegraphics[width=15cm]{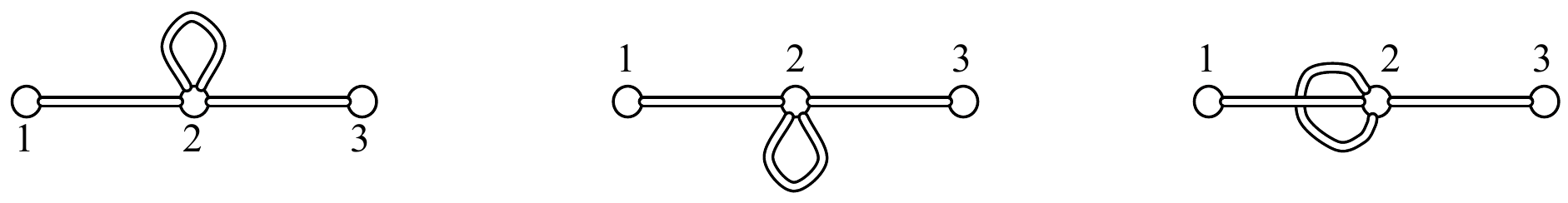}
\caption{Graphs forming the set $G_{0,3}(1,4,1)\,.$}
\label{fig:G03}
\end{figure}

\noindent As a less trivial example, consider $G_{0,3}(1,4,1)$ which consists of the graphs with three ordered vertices drawn on a sphere with degrees one, four, and one. There are three such graphs, as shown in Figure \ref{fig:G03}. 
 
\noindent Following notation of \cite{MotohicoCatalan} we let $D_{g,n}(\mu_1,\dots, \mu_n)$ stand for the number of marked graphs in the set $G_{g,n}(\mu_1, \dots, \mu_n)$ where each marked graph $\Gamma_v$ is counted with the weight $1/|{\rm Aut}_v (\Gamma_v)|$: 
\begin{equation}
\label{D_def}
D_{g,n}(\mu_1,\dots, \mu_n) = \sum_{\Gamma_v\in G_{g,n}(\mu_1, \dots, \mu_n)}\frac{1}{|{\rm Aut}_v (\Gamma_v)|}\,\,.
\end{equation}
\noindent For example, $D_{0,2}(2,2) = 1/2$ and $D_{0,1}(2) = 1/2$.  Due to Lemma \ref{lemma_vertical},  for $n>2$, the number $D_{g,n}(\mu_1,\dots, \mu_n)$ is an integer that counts distinct marked graphs in the set $G_{g,n}(\mu_1,\dots, \mu_n)$. For example, $D_{0,3}(1,4,1)=3$.

\noindent 
\noindent For every graph of the set $G_{g,n}(\mu_1,\dots, \mu_n)$, let us put an arrow on one of the $\mu_i$ half-edges incident to vertex $i$ for all $i=1,\dots, n$. Two such {\it decorated} marked graphs are identified if they are isomorphic as marked graphs and the marked isomorphism maps an arrowed half-edge to an arrowed one. We denote the resulting set of distinct decorated marked graphs by $\widehat G_{g,n}(\mu_1,\dots, \mu_n)$. The number of graphs in the set $\widehat G_{g,n}(\mu_1,\dots, \mu_n)$ is given by the following integer, see \cite{MotohicoCatalan}:
\begin{equation}
\label{DCgn}
C_{g,n}(\mu_1,\dots, \mu_n) =\mu_1\cdots \mu_n ~ \times D_{g,n}(\mu_1,\dots, \mu_n). 
\end{equation}
\noindent For example, $C_{0,2}(1,1)=1,\,$ $C_{0,2}(1,3)=3$, and   $C_{0,3}(1,1,4)=12$. As mentioned, all degrees $\mu_i$ should be positive integers with the  exception of a one-vertex graph. Thus for $n>1$ we have $C_{g,n}(\mu_1,\dots, \mu_n)=0$ if one of the $\mu_i$ is zero, however $C_{0,1}(0)=1.$ Note also that the sum of $\mu_i$ should be even for the corresponding $C_{g,n}$ to be nonzero. 

\begin{remark} In \cite{DumitrescuMulaseSafnukSorkin, MotohicoCatalan} the numbers $C_{g,n}$ were called \emph{generalized Catalan numbers} because in the case of one vertex,  $C_{0,1}(2m)$ is the $m$th Catalan number. 
In \cite{WalshLehman1},  the term \emph{dicings} was used for the decorated graphs from the set $\widehat G_{g,n}(\mu_1,\dots, \mu_n)$. 
\end{remark}
\vskip 1 cm

\noindent The precise relation between the coefficients $C_{g,n}$ and the enumeration of Feynman diagrams will be explained in Section \ref{sect_FD}.

\section{The numbers $C_{g,n}$ and the topological recursion}

\noindent The following recursion formula for  $C_{g,n}$ was derived in  \cite{WalshLehman1} and then rediscovered in \cite{DumitrescuMulaseSafnukSorkin, MotohicoCatalan} in the context of  the topological recursion:
\begin{multline}
\label{Cgn_recursion}
C_{g,n}(\mu_1,\dots, \mu_n) = \sum_{j=2}^n \mu_j C_{g,n-1}(\mu_1+\mu_j-2,\mu_2,\dots, \widehat{\mu_j}, \dots, \mu_n)
\\
+ \sum_{\alpha+\beta=\mu_1-2} \left[  C_{g-1,n+1}(\alpha, \beta, \mu_2,\dots, \mu_n) 
+ \sum_{\substack{g_1+g_2=g\\ I\sqcup J =\{2,\dots, n\}}}
 C_{g_1,|I|}(\alpha, \mu_I) C_{g_2,|J|}(\beta, \mu_J)
   \right],
\end{multline}
\noindent where we use the notation from \cite{DumitrescuMulaseSafnukSorkin, MotohicoCatalan}, namely $|I|$ denotes the number of elements in the set $I$, the symbol $\sqcup$ stands for disjoint union, $\mu_I=(\mu_i)_{i\in I}$, and the hat symbol marks the omitted argument.

\noindent In \cite{DumitrescuMulaseSafnukSorkin, MotohicoCatalan}, this recursion formula was shown to be related by a Laplace transform to the topological recursion of Eynard-Orantin. Here we describe this relationship.

\noindent
Consider the following infinite hierarchy of differentials defined, for $(g,n)$ such that $2g-2+n>0$, in terms of the $C_{g,n}$ by
\begin{equation}
\label{Wgnc}
W^C_{g,n}(t_1,\dots, t_n) = (-1)^n \sum_{(\mu_1,\dots, \mu_n)\in\mathbb Z_+^n} C_{g,n}(\mu_1,\dots, \mu_n) {\rm e}^{-\langle w, \mu\rangle}dw_1\cdots dw_n\,\,, 
\end{equation}
where $w=(w_1, \dots, w_n), \quad \mu=(\mu_1,\dots, \mu_n)$, the scalar product is defined by $\langle w,\mu \rangle = \sum \mu_i w_i,$ and the variables $t_i$  are related to $w_i$ by 
\begin{equation}
{\rm e}^{w_i} = \frac{t_i+1}{t_i-1}+\frac{t_i-1}{t_i+1} \qquad\mbox{for}\quad i=1,2,\dots, n.  \label{ttow}
\end{equation}

\noindent It was shown in  \cite{DumitrescuMulaseSafnukSorkin, MotohicoCatalan}  that differentials $W^C_{g,n}$ \eqref{Wgn} with $(g,n)$ such that $2g-2+n>0$ except $W^C_{1,1}$   satisfy the Eynard-Orantin topological recursion on the curve defined by the equation
\begin{equation}
\label{CatalanCurve}
\tilde{x}+\tilde{y}+\frac{1}{\tilde{y}}=0\,.
\end{equation}
\noindent For this curve, one considers two parameters $   \tilde{ y}$ and $t$ away from the point at infinity $( \tilde{ x}, \tilde{ y})=(\infty, \infty)$ related to each other by 
\begin{equation*}
 \tilde{ y}=\frac{1+t}{1-t}\, , 
\end{equation*}
which from Eq.(\ref{CatalanCurve}) gives
\be \tilde{x} =  2  \, \frac{t^2+1}{t^2-1} \, .  \nonumber 
\ee
\noindent The Galois involution in terms of these parameters becomes $ \tilde{ y}^*=1/ \tilde{ y}$ and $t^*=-t$ and the simple ramification points are located at $ \tilde{ y}=\pm 1$ or $t=0,\infty$.

\noindent However, to obtain differentials \eqref{Wgnc} by recursion on the curve \eqref{CatalanCurve} the definition of one of the initial differentials $W_{0,2}$ is modified in \cite{DumitrescuMulaseSafnukSorkin, MotohicoCatalan} as follows: 
\begin{equation}
\label{modified_w02}
W_{0,2}^C(t_1,t_2) =\frac{dt_1\cdot dt_2}{(t_1-t_2)^2} - \frac{d \tilde{ x}_1\cdot d \tilde{ x}_2}{( \tilde{ x}_1- \tilde{ x}_2)^2} = \frac{dt_1\cdot dt_2}{(t_1+t_2)^2}\,.
\end{equation}
\noindent With this $W_{0,2}$ and with the standard $W^C_{0,1}= \tilde{ y}d \tilde{ x} = 8t(t+1)dt/(t^2-1)^3$, the recursion kernel according to the above definition is
\begin{equation}
\label{CatalanKernel}
K^C(t,t_1)= -\frac{1}{64} \left( \frac{1}{t+t_1} + \frac{1}{t-t_1}  \right) \frac{(t^2-1)^3}{t^2}\frac{dt_1}{dt}\,.
\end{equation}

\noindent It turns out, see \cite{HO}, that all ``stable'' ($2g-2+n>0$) differentials $W^C_{g,n}$ produced by this recursion on the curve \eqref{CatalanCurve} can be obtained by recursion \eqref{recursion}-\eqref{K} starting with the following harmonic oscillator curve from the family \eqref{HOfamily}:% so-called {\it harmonic oscillator curve}, the algebraic curve defined by the equation
\begin{equation}
\label{HOcurve}
y^2=x^2-2\;.
\end{equation}
\noindent Moreover, the recursion on the curve \eqref{HOcurve} also produces $W^C_{1,1}$ which required a special treatment in \cite{DumitrescuMulaseSafnukSorkin, MotohicoCatalan}. The next theorem is a corollary of results from \cite{DumitrescuMulaseSafnukSorkin, MotohicoCatalan} and \cite{HO}.

\begin{theorem}
\label{thm_recursion}
Differentials $W^C_{g,n}$ \eqref{Wgnc}  satisfy the Eynard-Orantin topological recursion on the harmonic oscillator curve \eqref{HOcurve}. 
\end{theorem}

\noindent {\it Proof.} Note that the following change of variables
\begin{equation*}
{x}=-c\frac{\tilde{x}}{2},\qquad \qquad {y}=c\left(\tilde{y}+\frac{\tilde{x}}{2}\right)
\end{equation*}
\noindent transforms the  curve \eqref{HOfamily} into the curve \eqref{CatalanCurve} $\tilde{x}+\tilde{y}+{1}/{\tilde{y}}=0\,,$ which implies the following relation on the respective local parameters: $t=- \epsilon z.$  With this identification, the recursion kernel $K^H$ coincides with the kernel \eqref{CatalanKernel} from \cite{DumitrescuMulaseSafnukSorkin} if we put $c^2=2\,,$ that is if we specialize to the curve \eqref{HOcurve}.

\noindent Given that the initial differential $W_{0,1}$ does not enter the recursion formula \eqref{recursion} (it is only used for obtaining the recursion kernel $K$), the only difference in running the recursion for the curves \eqref{CatalanCurve} and \eqref{HOcurve} is due to the difference between the initial differentials $W_{0,2}^C$ and $W_{0,2}^H\,$. However, a careful examination of the recursion formula \eqref{recursion} shows that starting at the indices $(g,n)$ satisfying $2g-2+n\geq 2$ in the left hand side, the differential $W_{0,2}$ only enters in the following combination: 
$W_{0,2}(q^*,p_k) - W_{0,2}(q,p_k)$. This quantity is obviously the same for $W_{0,2}=W_{0,2}^H(z_1,z_2)$ and $W_{0,2}=W_{0,2}^C(t_1,t_2)$.  Therefore, starting from the second generation of $W_{g,n}$, that is from the indices $(g,n)$ such that $2g-2+n\geq 2$, the recursion formula \eqref{recursion} applied to the harmonic oscillator curve and the recursion from \cite{DumitrescuMulaseSafnukSorkin} coincide.

\noindent It can be shown by direct computation, see \cite{HO}, that the recursion for the curve \eqref{HOcurve} produces $W^C_{1,1}$ and $W^C_{0,3}$ from \cite{DumitrescuMulaseSafnukSorkin}. Thus $W^H_{g,n}(z_1,\dots, z_n)=W^C_{g,n}(t_1,\dots, t_n)$ with $t=- \epsilon z$ and $c^2=2\;.$
$\Box$

\subsection{Obtaining the $C_{g,n}$ from the $W^H_{g,n}$}

\noindent Once the $W^H_{g,n}(z_1, \ldots, z_n)$ have been obtained by topological recursion, it is simple to extract the $C_{g,n} (\mu_1, \ldots, \mu_n)$ using
\begin{equation}
\label{Wgn}
W^H_{g,n}(t_1,\dots, t_n) = (-1)^n \sum_{(\mu_1,\dots, \mu_n)\in\mathbb Z_+^n} C_{g,n}(\mu_1,\dots, \mu_n) {\rm e}^{-\langle w, \mu\rangle}dw_1\cdots dw_n\,\, .
\end{equation}

\noindent From now on we will use the identifications from the proof of Theorem \ref{thm_recursion}, namely 
\begin{equation*}
t=-\epsilon z \qquad\mbox{and}\qquad c^2=2,
\end{equation*}
\noindent and write $W^H_{g,n}(t_1, \ldots, t_n)$ for the differentials after these substitutions have been made. 
We first invert Eq.(\ref{ttow}), choosing the sign in order to be in agreement with  Eq.(\ref{Wgn}),
\be t_i = - \frac{   \sqrt{1+ 2 e^{-w_i}}}{\sqrt{1-2 e^{-w_i}}} \label{tiwi} \,.  \ee
\noindent Had we chosen  not to  include the factor $(-1)^n$ in Eq.(\ref{Wgn}), a positive sign would have been required in Eq.(\ref{tiwi}). 
  We will also need

\be dt_i =   \frac{2 e^{-w_i} }{(1-2e^{-w_i} )^2} \sqrt{\frac{1-2 e^{-w_i} }{1+2 e^{-w_i} }} \, dw_i  \, .\nonumber \ee

\noindent After expressing the  $W^H_{g,n}(t_1, \ldots, t_n)$   in terms of the variables $w_1, \ldots, w_n$,  we expand the results in powers of $e^{-w_i}$. According to Eq.(\ref{Wgn}),  the coefficient of $e^{-\langle w, \mu \rangle} \, dw_1 \ldots dw_n$, where $(\mu_1,\mu_2, \ldots, \mu_n) \in \mathbb{Z}_+^n $,  is  equal to $(-1)^n C_{g,n} (\mu_1, \ldots, \mu_n)$.

\noindent As an example, consider 
\be W^H_{1,1}(z_1) = \epsilon \frac{(z_1^2-1)^3}{2^6c^2 z_1^4} dz_1  
\qquad\mbox{ and thus }\qquad W^H_{1,1}(t_1) = - \frac{(t_1^2-1)^3}{2^6 c^2 t_1^4} dt_1  
\nonumber \, . \ee
\noindent Expressed in terms of $w_1$, this becomes
\be W^H_{1,1}(w_1) =    - \frac{2}{c^2} \frac{e^{-4w_1}}{(1-4 e^{-2w_1})^{5/2}}  dw_1  \,.
\nonumber
\ee
\noindent Taylor expanding, we obtain
\be
 W_{1,1}^H(w_1) =  -\frac{2}{3c^2}  
 \sum_{k=0}^\infty \frac{2^k \,  (2k+3)!! }{k!} ~e^{-(2k+4)w_1} \, dw_1\,. \nonumber
 \ee
\noindent From Eq.(\ref{Wgn}), this is equal to  $-\sum_{\mu_1=1}^\infty C_{1,1}(\mu_1) e^{-\mu_1 w_1} dw_1$. We therefore obtain
 \be C_{1,1} (2k+4) =   \frac{ 1}{3c^2 }  \frac{2^{k+1}}{k!}  (2k+3)!!  \, , ~~k=0,1 \ldots 
 \,. \label{c11} \ee

\noindent For calculating the numbers $C_{1,1}$, the power of the topological recursion approach is evident, as we have a closed form expression for  $C_{1,1}(\mu_1)$ instead of having to use the laborious recursion relation (\ref{Cgn_recursion}), which requires an ever increasing number of terms as $\mu_1$ increases.

\noindent Even though Eq.\eqref{Wgn} defines only ``stable'' ($2g-2+n>0$)  differentials by the Laplace transform, let us consider the ``unstable'' $W^H_{0,1}$ and $W^H_{0,2}$.  For  $W^H_{0,1}(t_1)$, we obtain in terms of $w_1$
% after setting $\epsilon c^2=-2$,

\be W^H_{0,1}(w_1)= - \frac{   \sqrt{1- 4 e^{-2 w_1}}  }{2 e^{-2w_1}}    dw_1
\, . \nonumber\ee  If we define $s_i:=e^{-w_i}$, what multiplies $d w_1$ is
\be -  \frac{   \sqrt{1-4 s_1^2}  }{2 s_1^2}  \nonumber \ee
\noindent which we recognize as the generating function of the Catalan numbers minus $1/(2s_1^2)$, see \cite{DumitrescuMulaseSafnukSorkin, MotohicoCatalan}.   To be precise,
\be -  \frac{   \sqrt{1-4 s_1^2}  }{2 s_1^2}  = \sum_{n=0} C_n  \, s_1^{2n} - \frac{1}{2 s_1^2} \, 
\ee
\noindent where the n-th Catalan number is given by 
\be C_n = \frac{1}{n+1} \binom{2n}{n} \, .\nonumber
\ee

\noindent Writing
\ba
W_{0,1}(w_1) &=:& - \frac{dw_1}{2 s_1^2} + \sum_{k=0}^\infty C_{0,1}(2k)  ~ s_1^{2k} ~ dw_1  
\nonumber \\
&=& 
 - \frac{dw_1}{2 e^{-2w_1}} + \sum_{k=0}^\infty C_{0,1}(2k)  ~ e^{-2k w_1} ~ dw_1\, ,
\nonumber
\ea
\noindent we obtain 
\ba C_{0,1} (2k) &=& C_k \nonumber \\
&=& 
\frac{1}{k+1} \binom{2k}{k} \, ,\label{c01}
\ea while $C_{0,1}(n)$ is zero when the argument is odd. 
\noindent Consider now $ W_{0,2}(t_1,t_2)$. It is given by 
\ba W_{0,2}(t_1,t_2) &=& \frac{dt_1 dt_2}{(t_1-t_2)^2}\nonumber \\
&=&   \ \frac{4 t_1 t_2}{(t_1^2-t_2^2)^2} \, dt_1 dt_2  + \frac{dt_1 dt_2}{(t_1+t_2)^2} \, .\nonumber \ea

\noindent As shown in \cite{DumitrescuMulaseSafnukSorkin}, the second term generates the $C_{0,2}(\mu_1,\mu_2)$. Going to the variables $w_1,w_2$ and using the notation $e^{-w_i}=s_i$, we find
\ba
W_{0,2}(w_1,w_2) &=& 
 \, \frac{ s_1 s_2}{(s_1-s_2)^2} ~ dw_1 dw_2\nonumber \\ 
  &~&  + \, \frac{4 s_1 s_2}{\sqrt{(1-4 s_1^2)(1-4 s_2^2)}} \frac{dw_1 dw_2}{\left( \sqrt{(1+2s_1)(1-2s_2)} + \sqrt{(1+2s_2)(1-2s_1)} \right)^2} 
\nonumber \\ 
&=&
\left( \frac{s_1 s_2}{(s_1-s_2)^2} +  \sum_{\mu_1,\mu_2=1}^\infty C_{0,2}(\mu_1,\mu_2) ~s_1^{\mu_1} s_2^ {\mu_2} \right) ~ dw_1 dw_2  \, ,
\nonumber
\ea
\noindent from which one can compute the $C_{0,2}(\mu_1,\mu_2)$.
Let us first introduce the function
\begin{equation} g(\mu) = 
2^{[\frac{\mu}{2}] } ~
  \frac{\left( 2 \left[ \frac{\mu-1}{2} \right]  +1 \right)!! }{ \left[\frac{\mu-1}{2}\right] !} \label{gmu}
\end{equation} where  $ [x] $ denotes the integer part of the real number $x$. 
\noindent In terms of this function, we find

\be C_{0,2} (\mu_1,\mu_2) =  \frac{\left(1+ (-1)^{\mu_1+\mu_2} \right) }{\mu_1+\mu_2}
  ~ g(\mu_1)  ~ g(\mu_2) \,,  \label{c02} \ee which agrees with
 \cite{DumitrescuMulaseSafnukSorkin}.

\noindent We find for $W_{0,3}$:
\begin{eqnarray*}
 W_{0,3}(w_1,w_2,w_3) &=& 
- \frac{ 1}{2}  \left(1 - \prod_{i=1}^3 \frac{1-2s_i}{1+2s_i} \right) 
\prod_{j=1}^3 \frac{s_j}{ (1-2s_j)^{3/2} \sqrt{1+2s_j}}  ~ dw_1 dw_2 dw_3 \nonumber \\
&=&- \frac{1}{2}  \prod_{i=1}^3 \left( \sum_{p_i,k_i=0}^\infty (-1)^{k_i} \frac{(1+2p_i)!! (2k_i-1)!!}{p_i! k_i!} s_i^{1+p_i+k_i} \right) 
~dw_1 \, dw_2 \, dw_3\\ &~& \quad + \frac{1}{2} \prod_{i=1}^3 \left( \sum_{p_i=0}^\infty \frac{ 2^{p_i}  (1+2p_i)!!  }{p_i!} (1-2s_i)s_i^{2p_i+1}
\right) ~dw_1 \, dw_2 \, dw_3 \\
&=& - 2 s_1^2 s_2 s_3 - 2 s_1 s_2^2 s_3- 2 s_1 s_2 s_3^3  - 12 s_1^4 s_2 s_3 - 12 s_1 s_2^4 s_3 - 12 s_1 s_2 s_3^4
\\ &~& ~~~~
  - 8 s_1^2 s_2^2 s_3^2 - 12 s_1^3 s_2^2 s_3  + \ldots\,,
\end{eqnarray*}
\noindent which, using once more Eq.(\ref{Wgn}),  is equal to
\begin{equation*}
 - \sum_{\mu_1,\mu_2,\mu_3=1}^\infty C_{0,3}(\mu_1,\mu_2,\mu_3) ~s_1^{\mu_1} \, s_2^{ \mu_2} \, s_3^{ \mu_3} ~dw_1 \, dw_2 \, dw_3 \, .
\end{equation*}
\noindent We can now obtain a closed form expression for the coefficients $C_{0,3}(\mu_1,\mu_2,\mu_3) $:

\begin{eqnarray}
C_{0,3}(\mu_1,\mu_2,\mu_3) 
%&=&- \frac{1}{2} \prod_{i=1}^3  g(\mu_i)  - \frac{1}{2} \prod_{i=1}^3
 % (-1)^{\mu_i} ~g(\mu_i)  \\
  &=&
   \frac{1}{2}  \Bigl(1 + (-1)^{\mu_1+ \mu_2 + \mu_3} \Bigr)  \prod_{i=1}^3  g(\mu_i)   \, , \label{c03}
\end{eqnarray}
\noindent where $g(\mu_i)$ was defined in Eq.(\ref{gmu}).

\noindent Proceeding similarly, one can obtain closed form formulas  for any given $C_{g,n}(\mu_1, \ldots, \mu_n)$. In the appendix the expressions for $C_{0,4}(\mu_1, \ldots, \mu_4)$ and $C_{1,2}(\mu_1,\mu_2)$ are given.   Calculating expressions for $W_{g,n}^H$ and $C_{g,n}$ for larger values of the indices poses no technical difficulties but they are quite lengthy.

\begin{remark}{\rm
For  a given $C_{g,n}(\mu_1, \ldots, \mu_n)$, the sum of the degrees $\mu_i$ satisfies an inequality that can be derived from the Euler relation (\ref{euler}).   Replacing $e$ by $\frac{1}{2}\sum \mu_i $ in Eq.(\ref{euler}), we obtain 
\begin{equation*} \sum_{i=1}^n \mu_i = 4g -4 + 2 f  + 2 n\, .  \end{equation*}
Clearly the sum of the  degrees must be even, as noted previously, but they satisfy an additional constraint coming from the fact that  the number of faces  is at least equal to one,  giving 
\be 
 \sum_{i=1}^n \mu_i \geq  2 (2g +n -1 ) \, . \label{ineq} \ee
 This shows  for example that $C_{1,1}(\mu)$ is nonzero at the condition that $\mu \geq 4$ as is made explicit in the general formula, Eq.(\ref{c11}).  The smallest degrees giving a nonzero value of $C_{0,4}(\mu_1,\ldots \mu_4)$ satisfy $\mu_1 + \mu_2 + \mu_3 + \mu_4 = 6$, in agreement with the expression given in the appendix. Two other examples that will be of use in Section  6.2 are  that $C_{1,2}(\mu_1,\mu_2)$ is nonzero only for $\mu_1 + \mu_2 \geq 6$  and $C_{1,3}(\mu_1,\mu_2,\mu_3)$ is nonzero at the  condition that $\mu_1+\mu_2+\mu_3 \geq 8$.
 }
 \end{remark}

\section{Rooted graphs}

\noindent It is convenient to introduce a root in a ribbon graph in order to remove the nontrivial   automorphisms of the graph.

\begin{definition}
\label{def_rooted}
A {\rm rooted  graph} is an unmarked ribbon graph with a distinguished half-edge, 
%$\widehat h$ 
the {\rm root} of the graph. The vertex incident to the root is called the \emph{root vertex}.
\end{definition}
\begin{remark}
A ribbon graph consisting of one vertex and no edges is considered to be a rooted graph. 
\end{remark}

\begin{definition}
An isomorphism, or a \emph{rooted isomorphism}, between two rooted graphs is an isomorphism between the ribbon graphs that maps the root to the root. 
\end{definition}

\noindent An isomorphism of a rooted graph to itself is called a {\it rooted automorphism} of the graph. 

\noindent For a given rooted graph, the group of its rooted automorphisms is trivial, see \cite{WalshLehman1}. 
 %Let us denote by $\Gamma^{(1)}$ a rooted graph whose underlying unmarked graph is $\Gamma$. 
 
\noindent In \cite{Feynman} the following generalization of the concept of a rooted graph was introduced. 
%\begin{definition}
%An {\rm $N$-rooted graph} is a ribbon graph with $N$ distinguished half-edges incident to $N$ distinct vertices labeled by $N$ %consecutive integers. The distinguished half-edges are called {\rm roots,} or {\rm root half-edges} and the corresponding vertices are %called {\rm root vertices}. 
%\end{definition}
%
\begin{definition}
\label{def_Nrooted}
An $N$-rooted graph is the data of a ribbon graph, $\Gamma = (H, \alpha, \sigma)$, with the choice of $N$ distinct ordered vertices, called the \emph{root vertices}, and the choice of $N$ half-edges, called the \emph{root half-edges}, or \emph{roots}, such that each root half-edge is incident to one of the root vertices. We call the \emph{$k$th root} the root half-edge incident to the $k$th root vertex. 
\end{definition}
\noindent In other words, an $N$-rooted graph is obtained from a ribbon graph by choosing $N$ distinct vertices, labeling them with numbers from $1$ to $N$, and at each of the chosen vertices placing an arrow on one of the half-edges incident to it. 
%Here is an alternative combinatorial definition of $N$-rooted graphs. 

\begin{definition}
\label{def_Nrooted_iso}
\noindent An isomorphism, or an \emph{$N$-rooted isomorphism}, between two $N$-rooted graphs is is an isomorphism between the underlying ribbon graphs that maps $k$th root to the $k$th root.  
\end{definition}

\noindent  Two isomorphic $N$-rooted graphs are identified. Similarly, an {\it $N$-rooted automorphism} of an $N$-rooted graph is an automorphism of the underlying ribbon graph which preserves the set of $N$ root vertices pointwise and maps roots to roots. Clearly, the only $N$-rooted automorphism of an $N$-rooted graph is the identity. \\

\subsection{Counting $N$-rooted graphs}
\label{sect_M2}

\noindent We have shown how to obtain the generalized Catalan numbers $C_{g,n}$ from the topological recursion on the harmonic oscillator curve \eqref{HOcurve}. Here we show that one can express the number of $N$-rooted maps in terms of the $C_{g,n}$. In the case $N=1$ we get the formula for one-rooted maps obtained in \cite{WalshLehman1}.

\noindent Let $\mu=(\mu_1,\dots, \mu_n)$ be a partition of an even integer $2e$ into $n$ strictly positive parts. 
It is convenient to regroup the parts $\mu_i$ into groups of equal values, that is we suppose that among the parts of $\mu$ there are $\m(\mu)$ distinct values with  $k_i$ copies of the value $\alpha_i$ for $i=1,\dots, \m(\mu)$. Clearly, for $\mu=(\mu_1,\dots, \mu_n)$ we have 
\begin{equation}
\sum_{i=1}^{\m(\mu)} k_i = n \qquad\mbox{and}\qquad \sum_{i=1}^{\m(\mu)} \alpha_i k_i = \sum_{j=1}^n \mu_j = 2e\;.
\label{sums}
\end{equation}

\noindent Given $e\in\mathbb N$, let us denote by $m_N(e)$ the number of distinct $N$-rooted graphs with $e$ edges.

\begin{theorem}
\label{thm_MN}
Let the numbers $C_{g,n}(\mu_1,\dots, \mu_n)$ be as defined in Section \ref{sect_Cgn}, and $e, N $ positive integers. %\in\mathbb N$ with $N\geq 1$. 
 The number $m_N(e)$ of  $N$-rooted graphs with $e$ edges is given by
% \sum_{\substack{\mu \,\vdash\, 2e\\\mu=(\mu_1,\, \mu_2, \ldots, \mu_n)}} 
\begin{equation}
\label{MN}
m_N(e) = ~\sum_{n=N}^{e+1} ~ \sum_{g=0}^{[\frac{1+e-n}{2}]} ~\sum_{\substack{\mu_1+\dots+\mu_n=2e \\ \mu_i \geq 1}} 
 \frac{\mu_1\cdots\mu_N}{(n-N)!}\frac{C_{g,n}(\mu_1,\dots, \mu_n)}{\mu_1\cdots\mu_n}\,,
\end{equation}
where $[\frac{1+e-n}{2}]$ denotes the integer part of the argument. Note that for a given choice of $N,$ the  minimum possible  value $e$ 
may take is $N-1$.
\end{theorem}

\noindent {\it Proof.} Using notation from Section \ref{sect_graphs}, let $\Gamma_v\in G_{g,n}(\mu_1,\dots, \mu_n)$ and let $\Gamma$ be its underlying unmarked graph. There are $k_1!\cdots k_{\m(\mu)}!$ ways to mark vertices of $\Gamma$ so that the resulting marked graph is in $G_{g,n}(\mu_1,\dots, \mu_n)$. Among these ways there might be equivalent ones - those giving rise to identical marked graphs. If there is a non-trivial unmarked automorphism $\phi\in{\rm Aut}(\Gamma)/{\rm Aut}_v(\Gamma)$, which acts  non-trivially on the set of vertices, then for every ordering of vertices $\Gamma_v$ there is an equivalent ordering given by $\phi(\Gamma_v)$. 
Therefore the number of distinct marked graphs $\Gamma_v\in G_{g,n}(\mu_1,\dots, \mu_n)$ having $\Gamma$ as the underlying unmarked graph is
\begin{equation}
\label{number}
\frac{k_1!\cdots k_{\m(\mu)}!}{|{\rm Aut}(\Gamma)/{\rm Aut}_v(\Gamma)|}\,.
\end{equation}
\noindent Let us denote by $D_{g,n}^\Gamma(\mu_1,\dots, \mu_n)$ the contribution to the number $D_{g,n}(\mu_1,\dots, \mu_n)$ of all the marked graphs whose underlying graph is $\Gamma$:
\begin{equation*}
D^\Gamma_{g,n}(\mu_1,\dots, \mu_n) = \sum_{\substack{\Gamma_v\in G_{g,n}(\mu_1, \dots, \mu_n)\\\Gamma \text{ is the underlying}\\\text{ unmarked graph for }\Gamma_v}}\frac{1}{|{\rm Aut}_v (\Gamma_v)|}\,\,.
\end{equation*}
\noindent Then we have 
\begin{equation*}
D_{g,n}(\mu_1,\dots, \mu_n)=\sum_\Gamma D_{g,n}^\Gamma(\mu_1,\dots, \mu_n)\,\,,
\end{equation*}
\noindent where the summation is over all distinct unmarked graphs $\Gamma$ for which there exists a marking $\Gamma_v\in G_{g,n}(\mu_1,\dots, \mu_n)$ having $\Gamma$ as the underlying unmarked graph. Using this notation and definition \eqref{D_def} of the number $D_{g,n}$, we deduce from Eq. \eqref{number}
\begin{equation}
\label{Dgn_contribution}
D_{g,n}^\Gamma(\mu_1,\dots, \mu_n) = \frac{k_1!\cdots k_{\m(\mu)}!}{|{\rm Aut}(\Gamma)|}\,\,.
\end{equation}

\noindent 
\noindent Now note that the third sum in the right hand side of \eqref{MN} can be seen as two nested sums: 
\begin{equation}
\sum_{\mu_1+\dots+\mu_n=2e}  = \sum_{\substack{\mu \,\vdash\, 2e\\\mu=(\mu_1,\, \mu_2, \ldots, \mu_n)}} \quad\sum_{\substack{\text{orderings}\\{ \text{of }  (\mu_1,\, \mu_2, \ldots, \mu_n)}}}, \label{nested}
\end{equation}
\noindent where the first sum in the right hand side is over all (unordered) partitions $\mu$ of $2e$ and the second sum is taken over the orderings of the $n$-tuple $(\mu_1, \mu_2, \ldots, \mu_n)$. Here an ordering is a multiset permutation of the $n$-tuple $(\mu_1, \mu_2, \ldots, \mu_n)$, that is a permutation  not distinguishing between repeated values. 
\noindent Let us use the following notation: denote by $\tau$ an ordering of $\mu=(\mu_1,\, \mu_2, \ldots, \mu_n)$ and denote the resulting $n$-tuple by $\tau(\mu) = (\mu_1^{(\tau)}, \mu^{(\tau)}_{2}, \dots, \mu^{(\tau)}_{n} )\,.$ 

\noindent Consider for example one partition of $6$ given by  $\mu=(\mu_1,\mu_2,\mu_3) =(1,1,4)$;  there are three different orderings of the triple: $\tau_1(\mu) = (\mu_1^{(\tau_1)},\mu^{(\tau_1)}_2,\mu^{(\tau_1)}_3)=(1,1,4)$, $\tau_2(\mu)=(\mu_1^{(\tau_2)},\mu^{(\tau_2)}_2,\mu^{(\tau_2)}_3)=(1,4,1)$, and $\tau_3(\mu)=(\mu_1^{(\tau_3)},\mu^{(\tau_3)}_2,\mu^{(\tau_3)}_3)=(4,1,1)$. 

\noindent Let us now fix $N\leq n$ and consider the following sum over all orderings of $\mu$
\be 
\sum_{\substack{\tau\,\, \in \text{ orderings}\\{ \text{of }  (\mu_1,\, \mu_2, \ldots, \mu_n)}}} \mu_{1}^{(\tau)}  \, \mu^{(\tau)}_{2} \cdots \mu_{N}^{(\tau)} \, , \nonumber
\ee 
\noindent  For the previous example of $(\mu_1,\mu_2,\mu_3) =(1,1,4)$, for which there are three orderings, let us consider the case $N=2$. Then the above sum over all orderings would give $1\cdot 1+ 1\cdot 4 + 4\cdot 1 = 9\;.$

\noindent This expression can be rewritten summing over all permutation of the $\mu_i$'s  treating the  repeated values as distinct. Denoting by $S_n$ the symmetric group of permutations, we have

\be
\sum_{\substack{\tau\,\, \in\text{  orderings}\\{ \text{of }  (\mu_1,\, \mu_2, \ldots, \mu_n)}}} \mu_{1}^{(\tau)}  \, \mu^{(\tau)}_{2} \cdots \mu_{N}^{(\tau)}
=
 \sum_{\substack{ \sigma\in S_n }}
 \frac{ \mu_{\sigma(1)} \, \mu_{\sigma(2)} \cdots \mu_{\sigma(N)}  
 %\,  \mu^0_{\sigma((N+1)_\tau)} \cdots \mu^0_{\sigma(n_\tau)} 
 }
 {k_1! k_2! \cdots k_{\m(\mu)}!}
 \,\,.\nonumber\ee

\noindent We can write this sum over all permutation of $n$ values as follows: 
\ba
 \frac{ (n-N)! }{k_1! k_2! \cdots k_{\m(\mu)}!}  \sum_{\substack{ \text{choice of}  \\ \text{$N$ terms out} \\ \text{of $(\mu_1,\dots, \mu_n)$} }}~~
   \sum_{\substack{ \gamma\in S_N }}
   ~~
 \mu_{\gamma(1)} \, \mu_{\gamma(2)} \cdots \mu_{\gamma(N)}  
 %\,  \mu^0_{\delta(N+1)} \cdots \mu^0_{\delta(n)}
\hskip 0.1cm, \nonumber 
 \ea 
\noindent where the factor $(n-N)!$ comes from the permutations of the remaining $(n-N)$ values of $\mu_i$ not entering the product in the numerator.

%\ba
 % \sum_{\substack{ \text{choice of}  \\ \text{$N$ terms out} \\ \text{of $(\mu_1,\dots, \mu_n)$} }}~~
 %  \sum_{\substack{ \gamma\in S_N }}
 % ~~
 %    \sum_{\substack{ \delta\in S_{n-N} }} \frac{ \mu_{\gamma(1)} \, \mu_{\gamma(2)} \cdots \mu_{\gamma(N)}  \,  \mu^0_{\delta(N+1)} \cdots \mu^0_{\delta(n)} }{k_1! k_2! \cdots k_{\m(\mu)}!}. \nonumber 
 %\ea 

% The sum over the  permutations $\delta$ of $n-N$ values gives trivially the factor  $(n-N)!$, so we have
%\be 
 %\sum_{\substack{ \sigma\in S_n }}
 %\frac{ \mu_{\sigma(1)} \, \mu_{\sigma(2)} \cdots \mu_{\sigma(N)}  \,  \mu^0_{\sigma(N+1)} \cdots \mu^0_{\sigma(n)} }{k_1! k_2! \cdots k_{\m(\mu)}!}
% = 
% \frac{(n-N)!}{k_1! k_2! \cdots k_{\m(\mu)}!} 
%  \sum_{\substack{ \text{choice of}  \\ \text{$N$ terms out}\\ \text{of $(\mu_1,\dots, \mu_n)$} }}~~
%   \sum_{\substack{  \gamma\in S_N  }}   \mu_{\gamma(1)} \, \mu_{\gamma(2)} \cdots \mu_{\gamma(N)} \,.
% \ee
\noindent The last sums give the number of ways to choose $N$ ordered roots in the graph $\Gamma$ disregarding the fact that there might be choices producing equivalent $N$-rooted graphs. 
\noindent  Introducing notation 
 $R^\Gamma_N$ for the number of choices of an ordered subset of $N$ half-edges in a given graph $\Gamma$, we obtain 
%  we finally obtain
% \be 
 %\sum_{\text{orderings}} \mu_1  \, \mu_2 \ldots \mu_N = 
 %\frac{(n-N)!}{k_1! k_2! \ldots k_m!} \times  ( \text{ number of $N$-rooted maps for a  given set of degrees }\mu_1, \mu_2 \ldots \mu_n )\;.
 %\ee

 \be 
 \label{L}
\sum_{\substack{\tau\,\, \in \text{  orderings}\\{ \text{of }  (\mu_1,\, \mu_2, \ldots, \mu_n)}}} \mu_{1}^{(\tau)}  \, \mu^{(\tau)}_{2} \cdots \mu_{N}^{(\tau)} = 
 \frac{(n-N)!}{k_1! k_2! \cdots k_{\m(\mu)}!} \times  R_N^\Gamma \,\,.
 \ee

\noindent  
 %Let $\Gamma_v\in\Gamma_{g,n}(\mu_1,\dots, \mu_n)$ and let $\Gamma$ be its underlying unmarked graph with $e$ edges. 
\noindent  Denote by $m_N^\Gamma(e)$ the contribution to the number $m_N(e)$ of all the $N$-rooted graphs whose underlying unrooted unmarked graph is $\Gamma$.  Every non-trivial automorphism $\phi\in{\rm Aut}(\Gamma)$ allows for identification of  $N$-rooted graphs included in the number $R_N^\Gamma$. Therefore, the contribution of $\Gamma$ to $m_N(e)$ is
\begin{equation*}
m_N^\Gamma(e)=\frac{R^\Gamma_N}{|{\rm Aut}(\Gamma)|}\,\,. 
\end{equation*}

\noindent On the other hand, obtaining $|{\rm Aut}(\Gamma)|$  from \eqref{Dgn_contribution} and $R_N^\Gamma$ from \eqref{L}, we have
\begin{equation*}
m_N^\Gamma(e)= \frac{D_{g,n}^\Gamma(\mu_1,\dots, \mu_n)}{(n-N)!} \sum_{\substack{\tau\,\, \in \text{  orderings}\\{ \text{of }  (\mu_1,\, \mu_2, \ldots, \mu_n)}}} \mu_{1}^{(\tau)}  \, \mu^{(\tau)}_{2} \cdots \mu_{N}^{(\tau)}  \,\,.
\end{equation*}

\noindent Summing this over all distinct unmarked graphs $\Gamma$ for which there exists a marking $\Gamma_v\in G_{g,n}(\mu_1,\dots, \mu_n)$ having $\Gamma$ as the underlying unmarked graph, we see that the contribution to $m_N(e)$ that comes from a given partition $\mu$ of $2e$ is given by 
\begin{equation}
\label{temp}
\frac{D_{g,n}(\mu_1,\dots, \mu_n)}{(n-N)!}\sum_{\substack{\tau \,\, \in \text{  orderings}\\{ \text{of }  (\mu_1,\, \mu_2, \ldots, \mu_n)}}} \mu_{1}^{(\tau)}  \, \mu^{(\tau)}_{2} \cdots \mu_{N}^{(\tau)}  \,\,.\end{equation}
\noindent Now, using \eqref{DCgn} and summing \eqref{temp} over all unordered partitions $(\mu_1,\dots, \mu_n)$ of $2e$ with $n$ positive parts, we obtain the contribution to $m_N(e)$ of all the graphs of genus $g$ with $n$ vertices. Summing further over all possible values of $g$ and $n$ and determining the summation limits using Eqs.(\ref{euler}) and (\ref{ineq}), we obtain the statement of the theorem.
$\Box$

\subsection{One-rooted graphs}
\label{sect_M1}

\noindent The following formula for the number $m_1(e,g)$ of rooted maps (that is maps with a distinguished half-edge) of genus $g$ with $e$ edges was derived in \cite{WalshLehman1}: 
\begin{equation}
\label{Intro_WalshLehman}
m_1(e,g)= \sum_{n=1}^\infty  ~ \frac{2e}{n!}  \sum_{\mu_1+\mu_2+\dots+\mu_n=2e}  \frac{C_{g,n}(\mu_1,\dots, \mu_n)}{\mu_1\dots\mu_n}\;.
\end{equation}

\noindent Our theorem, Eq.(\ref{MN}), predicts for this quantity
\be 
m_1(e,g)= \sum_{n=1}^\infty  ~ \sum_{\mu_1+\mu_2+\dots+\mu_n=2e}  \frac{\mu_1}{(n-1)!} 
\frac{C_{g,n}(\mu_1,\dots, \mu_n)}{\mu_1\dots\mu_n}\;. \label{ours}
\ee

\noindent To prove the equivalence of these two expressions, let us replace the sum in Eq.(\ref{ours}) by two nested sums over partitions and over orderings, as in Eq.(\ref{nested}).  Since the $C_{g,n}(\mu_1,\dots, \mu_n) $ do not depend on the ordering of the arguments, we may rewrite Eq.(\ref{ours}) as 
\begin{equation}
m_1(e,g)=  \frac{1}{(n-1)!} \sum_{n=1}^\infty  ~ \sum_{\substack{\mu \,\vdash\, 2e\\\mu=(\mu_1,\, \mu_2, \ldots, \mu_n)}} \quad \frac{C_{g,n}(\mu_1,\dots, \mu_n)}{\mu_1\dots\mu_n} 
\quad \sum_{\substack{\tau\,\, \in \text{ orderings}\\{ \text{of }  (\mu_1,\, \mu_2, \ldots, \mu_n)}}} \mu_{1}^{(\tau)} 
\;. \label{cjjd}
\end{equation}
\noindent Recall that  $\mu_1^{(\tau)}$ may take $\rho(\mu)$ distinct values denoted $\alpha_1, \ldots ,\alpha_{\rho(\mu)}$ and that each of these values appear a number of $k_1 , \ldots, k_{\rho(\mu)}$ times, respectively. The  sum over orderings is then given by 
\ba
\sum_{\substack{\tau\,\, \in \text{ orderings}\\{ \text{of }  (\mu_1,\, \mu_2, \ldots, \mu_n)}}} \mu_{1}^{(\tau)} 
&=& \alpha_1~ \frac{(n-1)!}{(k_1-1)!~ k_2!~ \ldots k_{\rho(\mu)}!} + \ldots + \alpha_\rho(\mu)~ \frac{(n-1)!}{k_1! k_2~ \ldots (k_{\rho(\mu)}-1)!}    \nonumber  \\
&=& \frac{(n-1)!}{k_1! k_2 ! \ldots k_{\rho(\mu)}!} \sum_{i=1}^{\rho(\mu)} k_i \alpha_i \nonumber \\
&=& 2e \frac{(n-1)! }{k_1! k_2! \ldots k_{\rho(\mu)}!} \, , \nonumber
\ea 
\noindent where in the last step we have used  Eq.(\ref{sums}). Using this result in Eq.(\ref{cjjd}) we obtain 
\begin{equation}
m_1(e,g)=  \sum_{n=1}^\infty  ~ \sum_{\substack{\mu \,\vdash\, 2e\\\mu=(\mu_1,\, \mu_2, \ldots, \mu_n)}} \quad \frac{2e}{k_1!  k_2 !\ldots k_{\rho(\mu)}!}  \frac{C_{g,n}(\mu_1,\dots, \mu_n)}{\mu_1\dots\mu_n} \, .  \nonumber
\end{equation}
\noindent Using now 
\begin{equation}
\sum_{\substack{\tau\,\, \in \text{ orderings}\\{ \text{of }  (\mu_1,\, \mu_2, \ldots, \mu_n)}}}  1 = \frac{n!}{k_1! k_2! \ldots k_{\rho(\mu)}!} \,\, ,  \nonumber
\end{equation}
\noindent we finally have
\begin{eqnarray}
m_1(e,g) &=&   \sum_{n=1}^\infty  ~ \sum_{\substack{\mu \,\vdash\, 2e\\\mu=(\mu_1,\, \mu_2, \ldots, \mu_n)}} \quad 
\sum_{\substack{\tau\,\, \in \text{ orderings}\\{ \text{of }  (\mu_1,\, \mu_2, \ldots, \mu_n)}}}  
\frac{2e}{n!}  \frac{C_{g,n}(\mu_1,\dots, \mu_n)}{\mu_1\dots\mu_n} \,  \nonumber 
\\ &=& 
 \sum_{n=1}^\infty  ~ \frac{2e}{n!}  \sum_{\mu_1+\mu_2+\dots+\mu_n=2e}  \frac{C_{g,n}(\mu_1,\dots, \mu_n)}{\mu_1\dots\mu_n}\, ,
 \nonumber
 \end{eqnarray}
\noindent which is Eq.(\ref{Intro_WalshLehman}).

\subsection{Counting Feynman diagrams}
\label{sect_FD}

\noindent We can now state the relation between the N-rooted graphs counted by $m_N(e)$ as given in Eq.(\ref{MN})  and Feynman diagrams. Our first result is that, as proven in \cite{Feynman}, $m_N(e)$ counts the number  of many-body Feynman diagrams, or   QED diagrams  when Furry's theorem is not  valid (due to the nature of the ground state) and, in particular, tadpoles are present.  However we will show below how Eq.(\ref{MN}) may be trivially modified to remove all tadpole diagrams or to enforce Furry's theorem.  The number $m_N(e)$ also counts the number of 
connected Feynman diagrams in the quantum field theory of a  two scalar fields, one real and one complex, with for only interaction the cubic term  $A \phi^\dagger \phi$.  It is using this latter quantum field theory that the number of Feynman diagrams was determined using the path integral approach in \cite{Feynman}.

\noindent In this correspondence, $N$ represents the number of external electron lines and $e$ is the number of internal photons lines (we consider diagrams with no external photon lines).  The order in the coupling constant is therefore simply $2e$.

\noindent In the sums of Eq.(\ref{MN}), the index $n$ counts the total number of electrons lines in the Feynman diagram, both internal and external. In the language of Feynman diagrams it is then obvious that, for a fixed number $e$ of photons, the smallest value of $n$ is $N$ (when all electron lines are external lines) and the maximum value of $n$ is  $e+1$ (which occurs when $N-1$ photons connect the external lines together and the remaining $e+1-N$ photons are part of  tadpoles so that there are $e+1-N$ fermion loops). 
 
\noindent The degrees $\mu_1, \ldots, \mu_n$ specify the number of photon lines connected to each of the fermion lines in the diagram.
 Since there are no external photon lines, we clearly have $\sum_{i=1}^n \mu_i = 2e$, which is twice the number of photon lines and the order in the coupling constant.

 \noindent We must warn against a possible source of confusion here: in the Euler  relation, Eq.(\ref{euler}), the number $n$ counts the number of vertices in the ribbon graph, which does not represent the number of vertices in the corresponding Feynman diagram but  the number of electron lines.  
 
\noindent To make this more clear, let us describe how to associate a ribbon graph to a Feynman diagram. To do so, one  must  draw 
all   photon lines connected to electron loops on the outside of the electron  loop.
In addition,   all photon lines connected to a given external electron line must be drawn on the same side of the electron line.

\noindent Then  the photon lines (but not the electron lines) are thickened  to turn them into ribbons and  the external electron  lines and electron loops are shrunk into  small disks which become the vertices of the corresponding ribbon graph. 
The number of vertices  in the ribbon graph is then equal to  the number of electron lines in the corresponding Feynman diagram and 
the number of ribbons connecting the vertices  is  the number of photon lines in the Feynman diagram. 

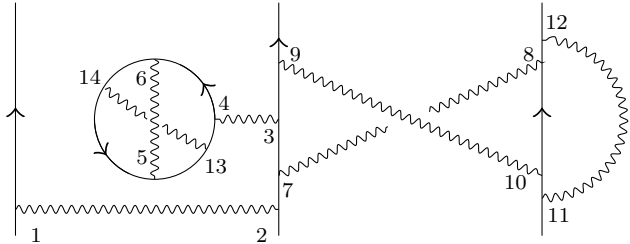
\begin{figure}
\centering
   \begin{center}
   \begin{tikzpicture}[node distance=1cm and 1cm]
\coordinate[] (v1);
\coordinate[below=1.55 cm of v1, label=right:$\,_1$] (v2);

\coordinate[above=1.55 cm of v1] (v3);
\coordinate[right=1.05 cm of v1] (v4);

\coordinate[right= 1.60 cm of v4] (v5);
\path (v5)+(350:0.10) coordinate[ label=above:$\,_4$];

\coordinate[right= 0.80  cm of v4] (center);
\coordinate[above= 0.80  cm of center] (top);
\coordinate[below= 0.80  cm of center] (bottom);
\coordinate[above=0.40 cm of v4](bv5);
\coordinate[right=0.14 cm of bv5](yaya);
\coordinate[above=0.40 cm of bottom](btop);
\coordinate[right=0.69 cm of btop](yoyo);
\coordinate[left=0.10 of center](yiyi);
\coordinate[right=0.10 of center](bon);
\coordinate[below=0.10 of bon](bill);
\path (bottom)+(110:0.55) coordinate[ label=below:$\,_5$];
\path (top)+(190:0.2) coordinate[ label=below:$\,_6$];
\draw[photonloop] (top) -- (bottom);
\draw[photonloop](yaya)--(yiyi);
\draw[photonloop](bill)--(yoyo);
\path (yoyo)+(0:0.10) coordinate[ label=below:$\,_{13}$];
\path (yaya)+(120:0.45) coordinate[ label=below:$\,_{14}$];
\draw[electronmiddle] (v2) -- (v3);
\coordinate[below=1.20 cm of v1] (v6);
\coordinate[right=3.50 cm of v6] (v22);
\semiloopt[electronf]{v4}{v5}{0};
\draw[photonloop] (v6) -- (v22);
\coordinate[below=0.35 cm of v22, label=left:$\,_2$] (v23);
\coordinate[above=3.10 cm of v23] (v24) ; 
\coordinate[right=3.50 cm of v23] (v33);
\coordinate[above=3.10 cm of v33] (v34) ;
\coordinate[above=1.55 cm of v33] (v31);
\coordinate[above=0.75 cm of v31](v35);

\coordinate[below=0.75 cm of v31](v36);
\coordinate[above=1.05 cm of v31](v37);
\path (v37)+(70:0.5) coordinate[ label=below:$\,_{12}$];
\coordinate[below=1.05 cm of v31](v38);
\coordinate[above=1.55 cm of v23](v21);
\path (v21)+(180:0.15) coordinate[ label=below:$\,_{3}$];
\coordinate[right=1.55 cm of v21](v39);
\coordinate[left=0.10 cm of v39](v40);
\coordinate[below=0.10 cm of v40](v41);
\coordinate[right=0.45 cm of v39](v42);
\coordinate[above=0.10 cm of v42](v43);
\draw[electronmiddle](v33)--(v34);
\draw[electronfff] (v23) -- (v24);

\draw[photonloop](v21) -- (v5);
\coordinate[below=0.75 cm of v21](v25);
\path (v25)+(15:0.1) coordinate[ label=below:$\,_7$];
\coordinate[above=0.75 cm of v21](v26);
\path (v26)+(60:0.4) coordinate[ label=below:$\,_{9}$];
\draw[photonloop](v25)--(v41);
\draw[photonloop](v43)--(v35);
\path (v35)+(120:0.4) coordinate[ label=below:$\,_{8}$];
\draw[photonloop](v26)--(v36);
\path (v36)+(160:0.4) coordinate[ label=below:$\,_{10}$];
\semiloopt[electronf]{v5}{v4}{180};
\semiloop[photonloop]{v37}{v38}{270};
\path (v38)+(0:0.2) coordinate[ label=below:$\,_{11}$];
\end{tikzpicture}
\caption{Feynman diagram corresponding to the permutations $\alpha=(1\,2)(3\,4)(5\,6)(7\,8)(9\,10)(11\,12)(13\,14)$, $\sigma=(\hat{1})(\hat{2}\,7\,3\,9) (\hat{11}\,10\,8\,12) (4\,6\, 14\,5 \, 13)$ on the set of half-edges $H=\{1, 2, \dots, 14\}$. 
\noindent The associated ribbon graph is shown in Figure \ref{ribbon_graph}. A different choice of labelling of half edges and the resulting permutations define an equivalent ribbon graph.} \label{fig:perms}
\end{center}   
   \end{figure}

\begin{figure}[htb]
\centering
\includegraphics[width=15cm]{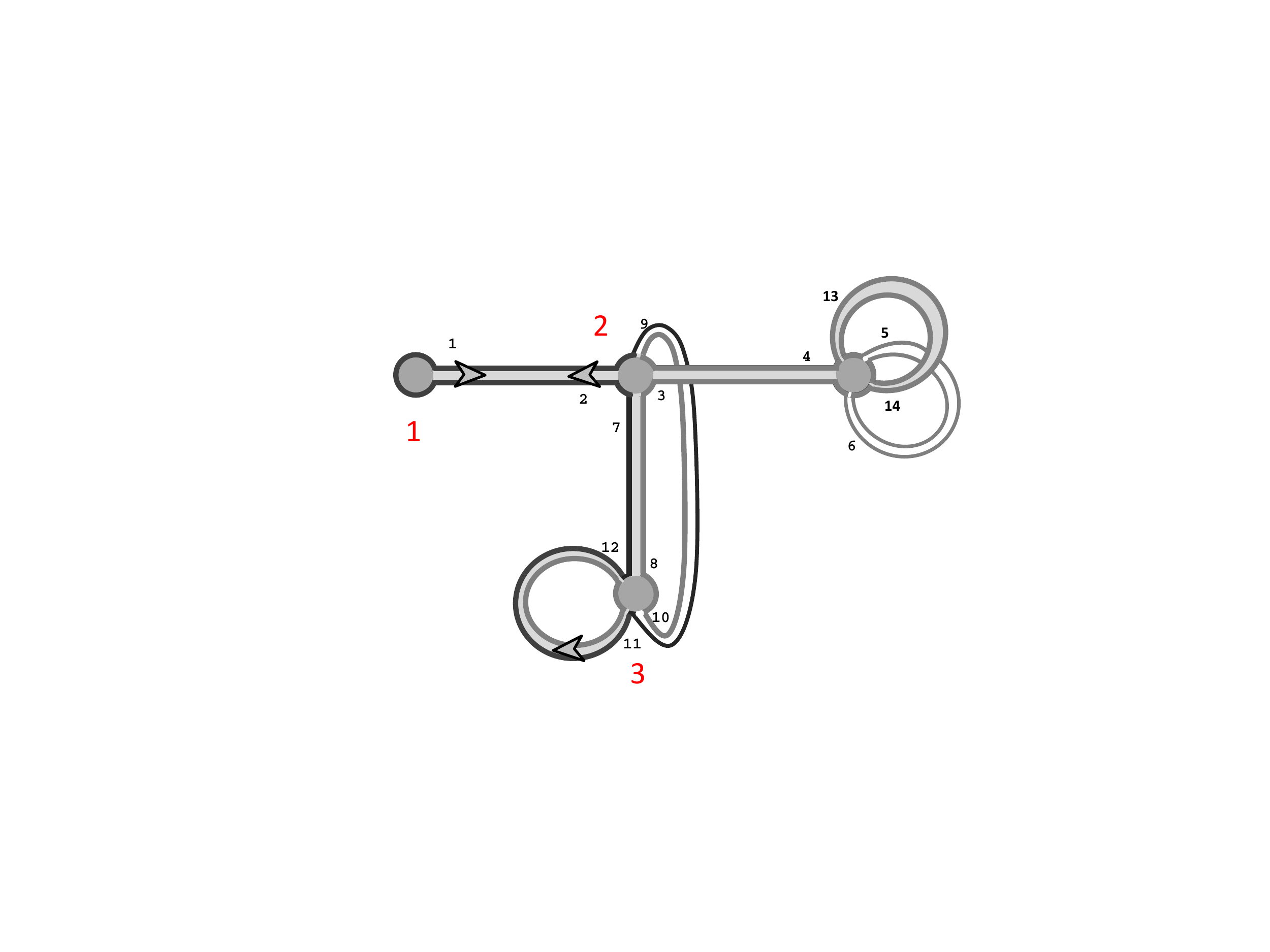}
\caption{The ribbon graph corresponding to Figure  \ref{fig:perms}. The three vertices with marked half edges correspond to the external electron lines and the labels 1 to 14 on the half edges  are the same as the ones shown in  Figure  \ref{fig:perms}.}
\label{ribbon_graph}
\end{figure}

\noindent As an example, a three point function Feynman diagram is shown in Figure  \ref{fig:perms} with its corresponding permutations and the corresponding ribbon graph is shown in Figure \ref{ribbon_graph}.

%The number of vertices and of ribbons connecting the vertices  in the ribbon graph are then equal to, respectively,  the number of electron lines and of photon lines in the corresponding Feynman diagram. 

  \begin{remark}
  {\rm
\noindent  The genus $g$ that appears in the coefficients $C_{g,n}$ refers to the genus of the corresponding ribbon graph, which is straightforward to determine.
  It is however  possible to determine the genus directly from the Feynman diagram without closing the external electron lines. The number of faces is  equal to the total number of closed paths necessary to cover both sides of all ribbons and all segments of electron lines connecting different ribbons. Note that the segments of the electron lines between the two sides of each ribbon are never covered. 
  Here it is understood that when one exits the Feynman diagram through one extremity of an external electron line, one then re-enters it through the second extremity.
  Once the number of faces is determined this way,  one uses Euler's formula, 
 \begin{equation} g= 1-f/2+e/2-n/2 \label{genus} \end{equation}
\noindent  to obtain the genus where, as already noted,  $e$  is the number of photon lines and $n$ is the number of electron lines. }
  \end{remark} 
  
\begin{figure}
\centering
   \begin{center}
   \begin{tikzpicture}[node distance=1cm and 1cm]
\coordinate[] (v1);
\coordinate[below=2.95 cm of v1] (v2);
\coordinate[above=3.25 cm of v1] (v3);
\coordinate[below=1.80 cm of v1] (v4);
\coordinate[below=1.20 cm of v1] (v5);
\coordinate[below=0.60 cm of v1] (v6);
\coordinate[above=0.30 cm of v1] (v7);
\coordinate[above=0.60 cm of v1] (v8);
\coordinate[above=2.00 cm of v1] (v9);
\coordinate[right=2.8 cm of v7] (v10);
\coordinate[above=1 cm of v10] (v11);
\draw[electronmiddle] (v2) -- (v3);
\semiloopq[photonloop]{v10}{v8}{270};
\semiloop[photonloop]{v6}{v4}{270};
\semiloop[photonloop]{v5}{v1}{90};
\coordinate[above=0.4 cm of v7] (v14);
\coordinate[right=0.6 cm of v14] (v12);
\coordinate[above=1 cm of v12] (v13);
\semiloopf[electronf]{v12}{v13}{180};
\coordinate[right=0.90 cm of v9] (v15);
\coordinate[below=4.25 cm of v3] (va);
\semiloopqq[photonloop]{va}{v15}{0};
%\path (v2)+(350:0.10) coordinate[ label=above:$\,_4$];
\end{tikzpicture}
\caption{Feynman diagram corresponding to the ribbon graph shown in Figure \ref{rib_fd}. } \label{self_energy}
\end{center}   
   \end{figure}
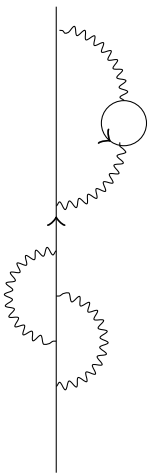

\begin{figure}[htb]
\centering
\includegraphics[width=15cm]{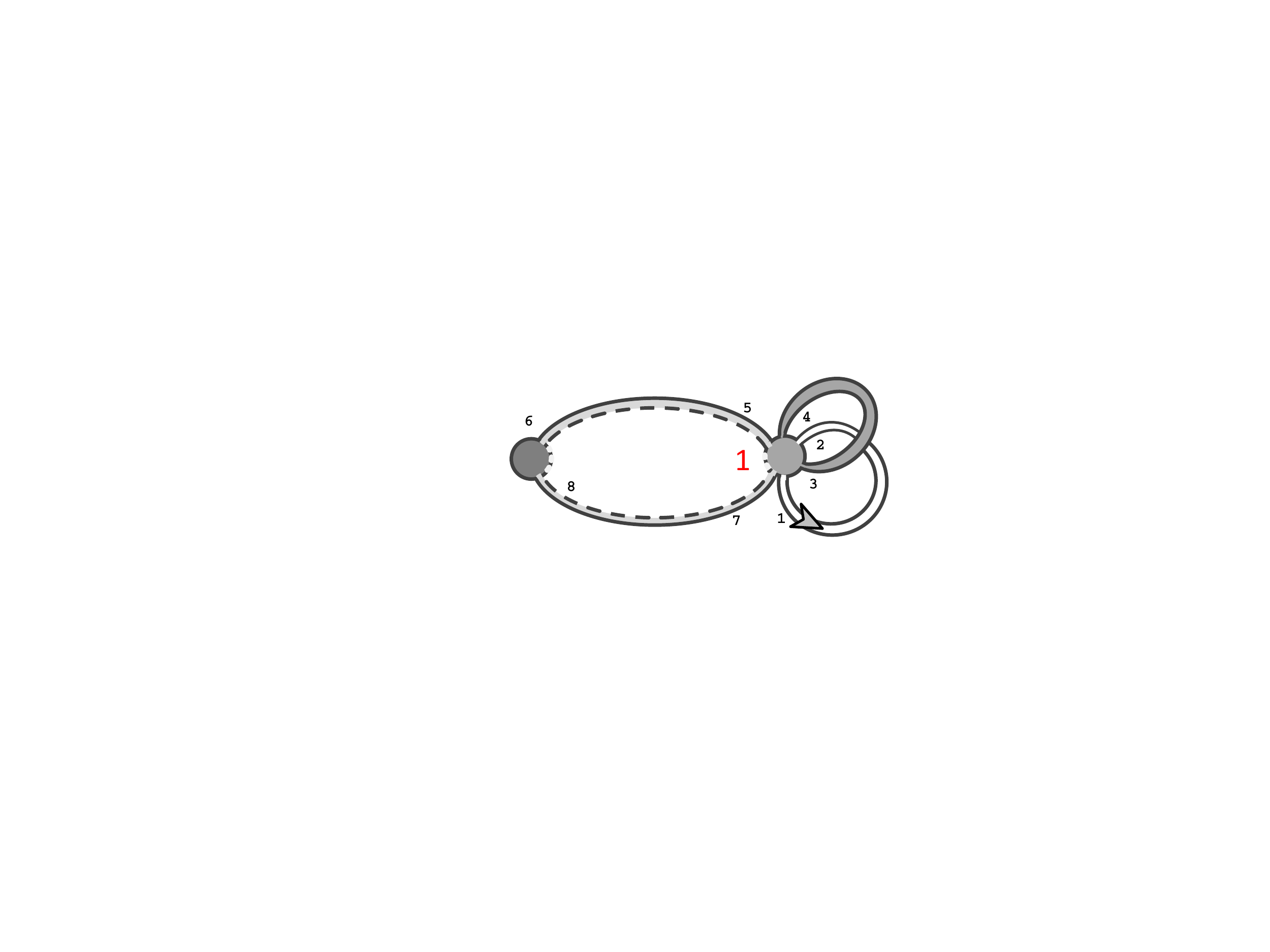}
\caption{The ribbon graph corresponding to Figure  \ref{self_energy}. It is specified by the permutations 
 $\alpha=(1\,2)(3\,4)(5\,6)(7\,8)$, $\sigma=(\hat{1}\,3\,2\,4\,5\,7)(6\,8) $ on the set of half-edges $H=\{1, 2, \dots, 8\}$.
The two paths required to cover the graph are shown as a dashed line and as a continuous line. This graph therefore has two faces.}
\label{rib_fd}
\end{figure}

  \begin{example} Consider the Feynman diagram of Figure \ref{self_energy}. The corresponding ribbon graph is shown in Figure \ref{rib_fd}  with the two paths (one made of a dashed line and the second shown as a continuous line) needed to cover the  graph. The number of faces is then two  and since $e=4$ and $n=2$, the genus of the ribbon graph corresponding to this Feynman diagram is equal to one. 
\end{example}

\begin{example}
As a second example, consider the ribbon graph of Figure \ref{ribbon_graph}. In that case, three paths are required to cover the graph so there are three faces. There are four vertices so  $n=4$ and there are seven edges so  $e=7$, giving  a genus $g=1$.
\end{example}

\vskip 0.5cm
\begin{remark}
{\rm
Since we are considering diagrams with no external photon lines and $N$ external fermion lines, there are $n-N$ fermion loops and the number of loops due to the photon propagators is $e+1 - n$, if $e$ is the number of photons.  We therefore obtain the following result: the total number of loops (in the sense of Feynman diagrams) is given by $e+1-N$. In other words, the number of $N$-rooted graphs with $e$ edges is equal to the number of Feynman diagrams with $e+1-N$ loops. 

This number is never negative since, for a given $N$, the smallest possible value of $e$ is $N-1$.
}
\end{remark}
\vskip 0.5cm

\noindent It is a simple matter to modify the formula (\ref{MN}) to count special types of Feynman diagrams. For example, it is trivial to exclude all diagrams with tadpoles by imposing that all closed electron loops must connect to more than one photon.  Recall that in Eq.(\ref{MN}), the  first $N$ indices refer to the external electron lines and the remaining indices, from $N+1 $ to $n$ refer to electron loops. To eliminate all tadpole diagrams, it is therefore only necessary to multiply the right hand side of Eq.(\ref{MN}) by a product of  Heaviside functions  forcing  the degrees of all the electron loops be at least equal to two:
\be H[\mu_{_{N+1}}-2 ] \, H[\mu_{_{N+2}}-2 ] \cdots H[\mu_{n}-2 ]  \,  \label{tad} \ee
where we use the convention $H[x] =1 $ for $x \geq 0$.

\noindent It is also easy to enforce Furry's theorem. This amounts to imposing that the degrees of all the electron loops be even, which is achieved by  simply multiplying Eq.(\ref{MN}) by 
\be\frac{1}{2^{n-N}} \left(1+(-1)^{\mu_{_{N+1}}} \right) \left(1+(-1)^{\mu_{_{N+2}}} \right) \ldots \left(1+(-1)^{\mu_{n}} \right) \, .
\label{furry}
\ee

\noindent Let us now show some examples of using Eq.(\ref{MN}). 

\noindent In the case of one external electron line, $N=1$, (so we are considering the corrections to the electron propagator), the number of  such diagrams as a function of the number of photon lines (which in this case gives also the number of loops) is well known \cite{cvitanovic} and given by 
\begin{equation} 1,  2, 10, 74, 706, 8\, 162, \, 110 \, 410  \ldots  \label{cvit}\end{equation}
\noindent  for $e=0,1,2, \ldots$.   The two diagrams for $e=1$ are the usual one-loop self-energy diagram plus the diagram with one tadpole.  Our formula for $m_1(e)$ reproduces trivially the results for $e=0$ and $e=1$.  Consider the next term in the series, corresponding to 
 two photon lines,  $e=2$.  The formula gives 
\begin{eqnarray} m_1(2) &=&
 ~\sum_{n=1}^{3} ~  \sum_{g=0}^{\left[ \frac{1+2-n}{2} \right] }~\sum_{\substack{\mu_1+\dots+\mu_n=4 \\ \mu_i \geq 1}} 
 \frac{\mu_1}{(n-1)!}\frac{C_{g,n}(\mu_1,\dots, \mu_n)}{\mu_1\cdots\mu_n}\, \nonumber \\
 &=&  C_{0,1}(4) +\frac{ C_{0,2}(1,3)}{3} + \frac{C_{0,2}(2,2)}{2} + C_{0,2}(3,1)  \nonumber \\
 &~& + \frac{1}{2} \left(  \frac{C_{0,3}(1,1,2)}{2} + \frac{C_{0,3}(1,2,1)}{2} + C_{0,3}(2,1,1) 
 \right) 
 % \nonumber \\ &~& +
+  C_{1,1}(4) 
   \, . \label{m12}
\end{eqnarray}
\noindent Using Eqs.(\ref{c11}), (\ref{c01}), (\ref{c02}) and (\ref{c03})   we obtain $m_1(2)=10$, in agreement with Eq.(\ref{cvit}).  

\noindent Each  term Eq.\eqref{m12} for $m_1(2)$ corresponds to a certain type of Feynman diagram.   Since we are considering  here one external electron line, $n-1$ is equal to  the number of electron loops. The degrees $\mu_i$ give the number of photon lines attached to the $i$-th electron  line, with $\mu_1$ being distinguished as the number of photons attached to the single external electron line.

\noindent For example consider the term 
\be \frac{1}{4} C_{0,3}(1,1,2) \, , \nonumber \ee
which corresponds, for $N=1$,  to  the diagram with one photon connecting  the external fermion line to a first  fermion loop, and a second photon line going from this first fermion loop to a second fermion loop, which is a tadpole. Clearly the term with $C_{0,3}(1,2,1)$ leads to the same diagram. Combining these two contributions and using $C_{0,3}(1,2,1)=2$ from Eq.(\ref{c03}), we find
\be 
\frac{1}{4} C_{0,3}(1,1,2)  + \frac{1}{4} C_{0,3}(1,2,1)  =1\, , \nonumber \ee
 \noindent which means that there is one such diagram. 

\noindent Consider now the term 
\be \frac{1}{2} C_{0,3}(2,1,1) = 1\, .\nonumber \ee
\noindent This corresponds to the  unique diagram with two  tadpoles attached to the external electron line.

\noindent The term $C_{0,2}(3,1) =3 $ in $m_1(2) $  counts the number of Feynman diagrams  with one  external electron line, one closed fermion loop,  three electron-photon vertices on the external line and only one photon attached to the fermion loop.  There are clearly three such diagrams corresponding to the tadpole attached before, between, or  after  the one loop self energy correction to the electron propagator.

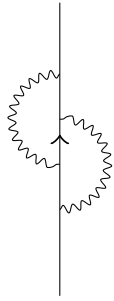
\begin{figure}
\centering
   \begin{center}
   \begin{tikzpicture}[node distance=1cm and 1cm]
\coordinate[] (v1);
\coordinate[below=2.95 cm of v1] (v2);
\coordinate[above=0.95 cm of v1] (v15);
\coordinate[above=2.95 cm of v1] (v3);
\coordinate[below=1.80 cm of v1] (v4);
\coordinate[below=1.20 cm of v1] (v5);
\coordinate[below=0.60 cm of v1] (v6);
\coordinate[above=0.30 cm of v1] (v7);
\coordinate[above=0.60 cm of v1] (v8);
\coordinate[above=2.00 cm of v1] (v9);
\coordinate[right=2.8 cm of v7] (v10);
\coordinate[above=1 cm of v10] (v11);
\draw[electronmiddle] (v2) -- (v15);
%\semiloopq[photonloop]{v10}{v8}{270};
\semiloop[photonloop]{v6}{v4}{270};
\semiloop[photonloop]{v5}{v1}{90};
\coordinate[above=0.4 cm of v7] (v14);
\coordinate[right=0.6 cm of v14] (v12);
\coordinate[above=1 cm of v12] (v13);
\end{tikzpicture}
\caption{Feynman diagram corresponding to the contribution of $C_{1,1}(4)$ in $m_1(2)$. The corresponding   ribbon graph shown in Figure \ref{Fig_torus}. } \label{two_loops}
\end{center}   
   \end{figure}

\noindent The term $C_{1,1}(4)$ in $m_1(2)$  corresponds to a Feynman diagram with  one electron line, $n=1$, one external electron line, $N=1$, and  two photons (since the number fo photons is equal to half the sum of the degrees) and whose ribbon graph is of genus one.  The ribbon graph is shown in  Figure \ref{Fig_torus} and the corresponding Feynman diagram is presented in Figure \ref{two_loops}. The ribbon graph has one face, one vertex and two edges, leading to $g=1$ according to Eq.(\ref{genus}).

\noindent The following generating function for $m_1(e)$ was found in \cite{ArquesBeraud} and rederived in \cite{Feynman} from a path integral approach:

\be
m_1(e) =  \sum_{k=0}^e (-1)^k 
 \sum_{\substack{a_1 + \ldots + a_{k+1}=e+1 \\ a_i \geq 1 } }  ~\prod_{j=1}^{k+1} (2 a_j-1)!!   \, .\nonumber
 \ee 
 \noindent This formula is more efficient than Eq.(\ref{MN})  to calculate the total number of Feynman diagrams for a corresponding number $e$ of internal photon lines, but Eq.(\ref{MN}) has the advantage of isolating the contributions from  each type of Feynman diagrams, {\it i.e.} according to the number of electron loops and the number of photon lines attached to each external electron line and to each electron loop. In particular, it is  possible to remove the tadpoles or to enforce Furry's theorem by using Eq.(\ref{MN}) multiplied by either Eq.(\ref{tad}) or Eq.(\ref{furry}).
 
\noindent For example, let's impose Furry's theorem to $m_1(2)$ as given by Eq.(\ref{m12}).   Multiplying by the factors of  Eq.(\ref{furry}) means that we drop all the $C_{g,n}( \mu_1, \ldots, \mu_2 )$ with $\mu_2, \ldots, \mu_n $ odd in Eq.(\ref{m12}). This leaves

\begin{eqnarray} m_1(2)\Big|_{(\text{Furry}) }
 &=&  C_{0,1}(4) + \frac{C_{0,2}(2,2)}{2} 
 % \nonumber \\ &~& +
+  C_{1,1}(4) \nonumber \\ &=& 4
   \, . \nonumber
\end{eqnarray}

\noindent Consider now $m_2(e)$, which begins at $e=1$ and corresponds to two external electrons. One finds 
\be
m_2(e) = 1,13,165, 2 \, 273, 34 \, 577 , 581 \, 133 , \ldots
\ee
\noindent for $e=1,2, \ldots$. 
Again, an explicit expression for $m_2(e)$ for arbitrary $e$ can be found in \cite{Feynman} but that expression is also not useful to isolate contributions from specific types of Feynman diagrams. In fact, \cite{Feynman} gives an algorithm to produce a closed form expression for $m_N(e)$ with arbitrary values of $N$ and $e$ but all these expressions do not separate the contributions from different classes of Feynman diagrams as Eq.(\ref{MN}) does.

\noindent Let us  reproduce the result $m_2(2) = 13$ using Eq.(\ref{MN}) which gives

\begin{eqnarray} m_2(2) &=&
 ~\sum_{n=2}^{3} ~  \sum_{g=0}^{\left[ \frac{3-n}{2} \right] }~\sum_{\substack{\mu_1+\dots+\mu_n=4 \\ \mu_i \geq 1}} 
 \frac{\mu_1 \, \mu_2}{(n-1)!}\frac{C_{g,n}(\mu_1,\dots, \mu_n)}{\mu_1\cdots\mu_n}\, \nonumber \\
 &=& \,  C_{0,2}(1,3) + C_{0,2}(3,1)+C_{0,2}(2,2) +  \,  C_{0,3}(2,1,1)+ C_{0,3}(1,2,1) + \frac{1}{2} \, C_{0,3}(1,1,2)
 \nonumber \\
 % \nonumber \\ &~& +
&=&  13 
  \nonumber  \, .
\end{eqnarray}

The presence of non trivial automorphisms in

\noindent Applying Furry's theorem here means discarding the terms with  three vertices for which  the degree of the third vertex is odd (the first two degrees giving the number of vertices on the two external electron lines).  This leaves
\begin{eqnarray*}
m_2(2)\Big|_{(\text{QED}) } &=& C_{0,2}(1,3) + C_{0,2}(3,1)+C_{0,2}(2,2)  + \frac{1}{2} \, C_{0,3}(1,1,2) \nonumber \\
&=&  9\, .
\end{eqnarray*}

\noindent As a more involved example, consider diagrams with three external electron lines, $N=3$ and $e \geq 2$.  The first few values  
are
\begin{equation*}
m_3(e) = 6,172,3\,834,81\,720,1\,775\,198\ldots \end{equation*}
for $e=2,3 \ldots $. 

  Equation (\ref{MN}) gives
\begin{eqnarray} m_3(2) &=&  3 \, C_{0,3}(1,1,2) \nonumber \\ &=& 6, \nonumber \\
    m_3(3) &=& 3 \, C_{0,3}(1,1,4) + C_{0,3} (2,2,2) + 6 \, C_{0,3}(1,2,3) \nonumber
    \\ &~&~~+ \frac{1}{3}  \, C_{0,4}(1,1,1,3) + 3 C_{0,4}(1,1,3,1) + \,  \frac{3}{2} C_{0,4}(1,1,2,2)+ 3  \, C_{0,4}(1,2,2,1) \nonumber \\
    &=& \, 172\, .
    \end{eqnarray}

\section{Expressing the WKB expansion in terms of  the coefficients $C_{g,n}$.}
\label{sect_WKBG}

\noindent We have now seen how applying topological recursion to the harmonic oscillator curve generates the multi-differentials $W^H_{g,n}$ which may be used  to both construct  the WKB expansion of the wave functions and  to calculate the coefficients $C_{g,n} $ which we used to  count QED Feynman diagrams. 

\noindent The formulas needed to obtain the WKB expansion are given in 
Eqs.(\ref{FgnHO}) and (\ref{keysn}), while the equations necessary to calculate the $C_{g,n}$ are Eqs.(\ref{Wgn})  and (\ref{tiwi}).  These two calculations imply that it is possible to express the WKB expansion in terms of the coefficients $C_{g,n}\,.$   For example, consider 
\ba S_2(x) &=&  
 \frac{1}{24c^2}  \frac{x^3-6c^2x}{(x^2-c^2)^{3/2}} \nonumber
 \\ &=:& \frac{1}{24 c^2} + \frac{1}{c^2} \sum_{k=1}  G(2k) \left( \frac{c}{2x} \right)^{2k} \, , \label{s2}
\ea
\noindent where $c$ is the constant that appeared in our elliptic curve and whose square we found to be $c^2=2$. We find it convenient to leave it as a parameter here to show the explicit dependence of $S_2$ on $c$.   The $x$ independent term in Eq.(\ref{s2}) is inconsequential as it contributes only an overall factor to  the wave function.  Using Eqs.(\ref{FgnHO}) and (\ref{keysn}), one finds 
\begin{eqnarray*} G(n) &:= & - \frac{2}{ 3!} \left( \sum_{\substack{i+j+k=n\\ i,j,k \geq 1}}
\frac{C_{0,3}(i,j,k)}{ i j k } + \frac{3}{2} 
\sum_{\substack{i+j=n\\ i,j\geq 1}}
 \frac{\widetilde{C}_{0,3}(i,j,0)}{i j }  + \frac{3}{4 n} \widetilde{C}_{0,3}(0,0,n) \right)  \nonumber \\
&~& \quad \quad \quad \quad    ~~ - 2 \,  \frac{C_{1,1}(n)}{n   } ~ H[n-4]\, , \end{eqnarray*}
\noindent with the coefficients $C_{1,1} $ and  $C_{0,3}$ given in Eqs.(\ref{c11}) and (\ref{c03}).

\noindent This expression requires some explanations.  As mentioned previously, the $C_{g,n} (\mu_1, \ldots, \mu_n)$ are taken to be zero whenever one of the degrees $\mu_i$ is zero, except for 
$C_{0,1}(0)=1$. In the above equation, the $\widetilde{C}_{0,3}$ are given by the same equation as the $C_{0,3}$, Eq.(\ref{c03}), except that we allow the degrees to be zero. 

\noindent This expression shows explicitly how a coefficient in the WKB expansion of the harmonic oscillator wavefunctions can be expressed in terms of the coefficients $C_{g,n}(\mu_1, \ldots, \mu_n)$ and we have seen how the same coefficients can be used to count Feynman diagrams in many body physics or in QED.
% The interpretation of the $\widetilde{C}_{0,3}(i,j,0)$ is that they represent the number of graphs with two marked vertices of degrees $i$ and $j$ and with one single point (corresponding to the zero) placed in any of the faces. Similarly,
% $\widetilde{C}_{0,3}(0,0,n)$  count the number of graphs with one marked point of degree $n$ with two points placed on any of the faces.

%This expression explicitly gives one term of the WKB expansion, $S_2(x)$ in terms of the numbers of Feynman diagrams in the cubic scalar field theory satisfying $2g+n-2 = 1$. 

\section{Appendix: $C_{0,4}(\mu_1 \mu_2, \mu_3, \mu_4)$ and $C_{1,2}(\mu_1,\mu_2)$.}
\noindent Here we give the expressions for the coefficients $C_{0,4}$ and $C_{1,2}$ without presenting the derivations. One finds
\be
C_{0,4} (\mu_1,\mu_2,\mu_3,\mu_4)= 
\begin{cases}
 \frac{\mu_1 +\mu_2 +\mu_3 +\mu_4-4}{2} ~g(\mu_1) \, g(\mu_2) \, g(\mu_3) \, g(\mu_4) 
~~~\text{ if all the } \mu_i \text{ are odd},\\ \, \\
 \frac{\mu_1 +\mu_2 +\mu_3 +\mu_4-2}{2} ~ g(\mu_1) \, g(\mu_2) \, g(\mu_3) \, g(\mu_4) 
~~~\text{ if all the  } \mu_i \text{ are even or if only two are even} , \\ \, \\ 0 ~~~\text{otherwise} ,
\end{cases} \nonumber 
\ee
\noindent where it is understood that $\mu_i > 0 $ and the function $g(\mu)$ is defined in Eq.(\ref{gmu}). Let us recall that  $W_{0,4} (s_1, s_2, s_3, s_4) $ is then given by 
\be W_{0,4}^H  (s_1, s_2, s_3, s_4) = \sum_{\mu_1, \mu_2, \mu_3, \mu_4 =1}^\infty  C_{0,4} (\mu_1,\mu_2,\mu_3,\mu_4)~ s_1^{\mu_1} s_2^{\mu_2} s_3^{\mu_3} s_4^{\mu_4} \,dw_1 \, dw_2 \,  dw_3 \,  dw_4  \nonumber \ee
\noindent where $s_i := e^{-w_i}$.

\noindent In order to present $C_{1,2}$, it proves convenient to first introduce the following function:
\be {\cal{F}}(\mu)  := \begin{cases}  2^{(\frac{\mu}{2})} \, \frac{(\mu+5)!!}{5 !! (\frac{\mu}{2})! }  ~~~\text{ if  } \mu \text{ is an even integer } \geq 0, \\ \, \\  0
~~~\text{otherwise} .
\end{cases} \nonumber 
\ee
\noindent The coefficients $C_{1,2}(\mu_1,\mu_2) $ are nonzero only if $\mu_1$ and $\mu_2$ have the same parity. When the two are even, one finds
\begin{eqnarray} C_{1,2}(\mu_1,\mu_2) &=&  4 {\cal{F}}(\mu_1-4) {\cal{F}}(\mu_2-2) \nonumber 
+4  {\cal{F}}(\mu_1-2) {\cal{F}}(\mu_2-4)+ 4 {\cal{F}}(\mu_1-6) {\cal{F}}(\mu_2-2) \nonumber  \\ &~&~~ +4{\cal{F}}(\mu_1-2){\cal{F}}(\mu_2-6)
-52 {\cal{F}}(\mu_1-4) {\cal{F}}(\mu_2-4)-16 {\cal{F}}(\mu_1-4){\cal{F}}(\mu_2-6) \nonumber   \\ &~&~~      -16 {\cal{F}}(\mu_1-6){\cal{F}}(\mu_2-4)
+320 {\cal{F}}(\mu_1-6) {\cal{F}}(\mu_2-6) \nonumber 
\, .
\end{eqnarray}   
\noindent Since  the function ${\cal{F}}$ vanishes when its argument is negative, we see that the first nonzero terms in $W^H_{1,2}$ have a sum of the exponents of $s_1 $ and $s_2$ equal to six, in agreement with Eq.(\ref{ineq}). Moreover, we see directly from the above that these terms are given by $4 s_1^4 s_2^2 + 4 s_1^2 s_2^4$.

\noindent When both arguments are odd, one obtains

\begin{eqnarray} C_{1,2}(\mu_1,\mu_2) &=&  5 {\cal{F}}(\mu_1-5) {\cal{F}}(\mu_2-1) \nonumber 
+5  {\cal{F}}(\mu_1-1) {\cal{F}}(\mu_2-5)+ 3 {\cal{F}}(\mu_1-3) {\cal{F}}(\mu_2-3) \nonumber  \\ &~&~~ -52{\cal{F}}(\mu_1-5){\cal{F}}(\mu_2-3)
-52 {\cal{F}}(\mu_1-3) {\cal{F}}(\mu_2-5)+208 {\cal{F}}(\mu_1-5){\cal{F}}(\mu_2-5) \nonumber   
\, ,
\end{eqnarray}  
\noindent which shows again that the terms of lowest exponents are $5 s_1^5 s_2 + 5 s_1 s_2^5  + 3 s_1^3 s_2^3$.

\noindent Let us write the first few terms of $W_{1,2}^H$ obtained using the above expressions:
\begin{eqnarray*} W_{1,2}^H (s_1,s_2) &=& 
 \sum_{\mu_1, \mu_2 =1}^\infty C_{1,2} (\mu_1,\mu_2) ~ s_1^{\mu_1} s_2^{\mu_2} \,dw_1 \, dw_2 \,  \nonumber \\ &=& dw_1 \, dw_2 \Bigl( 5 s_1^5 s_2 + 5 s_1 s_2^5 +4 s_1^4 s_2^2 + 4 s_1^2 s_2^4 + 3 s_1^3 s_2^3 + 70 s_1^7 s_2 + 70 s_1 s_2^7 + 60 s_1^6 s_2^2 + 60 s_1^2 s_2^6 
\nonumber \\ &~& ~~~+ 60 s_1^5 s_2^3 + 60 s_1^3 s_2^5 + 60 s_1^4 s_2^4 + 630 s_1^9 s_2 + 630 s_1 s_2^9 +  560 s_1^8 s_2^2 + 560 s_1^2 s_2^8 
\nonumber \\ &~& ~~~ + 630 s_1^7 s_2^3 + 630 s_1^3 s_2^7 + 600 s_1^6 s_2^4 + 600 s_1^4 s_2^6 + 600 s_1^5 s_2^5 + \ldots \Bigr) \, .
\end{eqnarray*}
\noindent One can verify that this indeed reproduces the Taylor expansion of $W_{1,2}$ given in Eq.(\ref{w12}) after making the change of variable
\begin{eqnarray*} z_i &=& - \epsilon \,  t_i \\
&=& t_i \\ &=& - \frac{\sqrt{1+ 2 e^{-w_i}}}{\sqrt{1-2 e^{- w_i}}} \\
&=& - \frac{\sqrt{1+ 2 s_i }}{\sqrt{1-2 s_i}}  \, .\end{eqnarray*}

\end{document}